\documentclass[aps,prx,reprint,superscriptaddress,nofootinbib,floatfix,preprintnumbers]{revtex4-2}

\usepackage{xcolor}
\usepackage{graphicx}
\usepackage{amsmath}
\usepackage{amsfonts}
\usepackage{amssymb}
\usepackage[normalem]{ulem}
\usepackage{physics}
\usepackage{float}
\usepackage{tikz}
\usepackage{subcaption}
\usetikzlibrary{arrows.meta, positioning, fit, backgrounds}
\usepackage{caption}
\usepackage{ragged2e}
\usepackage[normalem]{ulem}

\usepackage[colorlinks=true, linkcolor=blue, citecolor=blue, urlcolor=blue]{hyperref}

\newcommand{\lamet}{\texttt{lamet-agent}}

\begin{document}

\preprint{FERMILAB-PUB-26-0691-T}

\title{LaMET-Agent: An Agent Framework for Large-Momentum Effective Theory Analysis}

\thanks{Code available at \url{https://github.com/AI4LGT/lamet-agent}.}

\author{Jinchen He}
\email{jinchen@fnal.gov}
\affiliation{Fermi National Accelerator Laboratory, Batavia, IL 60510, USA}
\affiliation{Maryland Center for Fundamental Physics, University of Maryland, College Park, MD 20742, USA}
\affiliation{Physics Division, Argonne National Laboratory, Lemont, IL 60439, USA}

\author{Xiangyu Jiang}
\email{xj12@iu.edu}
\affiliation{Department of Physics, Indiana University, Bloomington, IN 47405, USA}

\author{Fei Yao}
\email{fyao@bnl.gov}
\affiliation{Physics Department, Brookhaven National Laboratory, Upton, New York 11973, USA}

\author{Dian-Jun Zhao}
\email{zhaodianjun@cuhk.edu.cn}
\affiliation{School of Science and Engineering, The Chinese University of Hong Kong, Shenzhen 518172, China}

\definecolor{softorange}{RGB}{232,190,145}
\definecolor{softblue}{RGB}{164,190,211}
\definecolor{softgreen}{RGB}{163,194,181}
\newcommand{\DJZ}[1]{{\color{brown}{DJZ:#1}}}
\newcommand{\JH}[1]{{\color{blue}JH: {#1}}} 
\newcommand{\FY}[1]{{\color{orange}FY: {#1}}} 
\newcommand{\JXY}[1]{{\color{green}JXY: {#1}}}

\begin{abstract}
Large-momentum effective theory (LaMET) provides a first-principles framework for computing the $x$ dependence of light-cone parton distributions from lattice QCD. 
Over the past decade, theoretical and numerical advances have established a mature multi-stage workflow for systematic calculation of parton physics, although its implementation still requires expert judgment and substantial repeated effort.
We present \lamet, an open-source large language model (LLM) agent framework that organizes this workflow into an executable, reproducible, and inspectable analysis pipeline.
The present release supports collinear quark distributions and implements correlator analysis, renormalization, Fourier transformation, perturbative matching, continuum, physical pion mass and infinite-momentum extrapolations, and automated result review.
We validate it on four end-to-end analyses: pion parton distribution functions in the gauge-invariant and Coulomb-gauge formulations, and pion and kaon distribution amplitudes, obtaining results consistent with the published calculations. Extensions to transverse-momentum-dependent distributions, generalized transverse-momentum-dependent distributions, and gluonic distribution functions are planned for subsequent releases.
\end{abstract}

\maketitle

\section{Introduction}
\label{sec:intro}

Over the past decade, lattice QCD calculations of parton physics have progressed rapidly toward a first-principles description of hadron structure in terms of quarks and gluons~\cite{Cichy:2018mum,Constantinou:2020hdm,Zhao:2025oto}. Parton distribution functions (PDFs) and distribution amplitudes (DAs) describe complementary aspects of one-dimensional longitudinal structure, while transverse-momentum-dependent distributions (TMDs) and generalized parton distributions (GPDs) extend this picture to transverse momentum and spatial structure. These intrinsically nonperturbative quantities are important both for understanding hadron mass and spin and for the experimental programs at Jefferson Lab~\cite{Dudek:2012vr}, RHIC~\cite{STAR:2014wox,PHENIX:2014gbf,PHENIX:2015fxo}, and the future EIC~\cite{Accardi:2012qut,AbdulKhalek:2021gbh}. Lattice QCD can provide first-principles calculations complementary to their phenomenological extraction from experimental data.

Obtaining the Bjorken-$x$ dependence of parton distributions from lattice QCD is nevertheless difficult: their definitions involve light-cone correlations, whereas classical lattice simulations evaluate Euclidean equal-time observables. The traditional local-operator approach accesses only the lowest few Mellin moments in practice, since higher moments encounter increasingly severe power-divergent operator mixing~\cite{Lin:2017snn}. Several complementary methods have been developed to access higher moments or the $x$ dependence~\cite{Liu:1993cv,Detmold:2005gg,Braun:2007wv,Davoudi:2012ya,Radyushkin:2017cyf,Ma:2017pxb,Chambers:2017dov,Shindler:2023xpd}. They provide valuable constraints, but reconstructing a full distribution from limited moments or short-distance correlations generally requires additional parameterization assumptions.

Large-momentum effective theory (LaMET)~\cite{Ji:2013dva,Ji:2014gla,Ji:2020ect,Ji:2024oka} provides an effective-field-theory route to the $x$ dependence, building on Feynman's observation that partons emerge as the infinite-momentum limit of static properties of a boosted hadron~\cite{Feynman:1969ej}. It evaluates time-independent spatial correlators, or quasi-distributions, in a hadron with finite but large momentum $P_z$ and relates them to light-cone distributions through perturbative matching and a systematic large-momentum expansion. In the moderate-$x$ region where these expansions are controlled, LaMET permits a systematically improvable point-by-point determination. LaMET has now been applied widely to PDFs, GPDs, TMDs, and light-cone wave functions~\cite{Xiong:2013bka,Lin:2014zya,Ji:2014hxa,Ji:2014lra,Sufian:2014jma,Ji:2015qla,Xiong:2015nua,Alexandrou:2016eyt,Chen:2016utp,Lin:2016qia,Yang:2016nfc,Yang:2016plb,Alexandrou:2017dzj,Alexandrou:2017huk,Alexandrou:2017qpu,Chen:2017mie,Chen:2017mzz,Constantinou:2017sej,Ishikawa:2017iym,Ji:2017oey,Ji:2017rah,Lin:2017ani,Stewart:2017tvs,Xiong:2017jtn,Zhang:2017bzy,Zhang:2017zfe,Alexandrou:2018eet,Alexandrou:2018pbm,Alexandrou:2018yuy,Chen:2018xof,Ebert:2018gzl,Fan:2018dxu,Izubuchi:2018srq,Ji:2018hvs,LatticeParton:2018gjr,Lin:2018pvv,Liu:2018hxv,Liu:2018tox,Zhang:2018diq,Zhang:2018nsy,Zhang:2018rls,Zhao:2018fyu,Alexandrou:2019dax,Alexandrou:2019lfo,Chai:2019rer,Chen:2019lcm,Constantinou:2019vyb,Ebert:2019tvc,Liu:2019urm,Shanahan:2019zcq,Wang:2019msf,Zhang:2019qiq,Alexandrou:2020qtt,Alexandrou:2020uyt,Alexandrou:2020zbe,Bhattacharya:2020cen,Bhattacharya:2020jfj,Chai:2020nxw,Chen:2020arf,Chen:2020iqi,Chen:2020ody,Ebert:2020gxr,Fan:2020nzz,Gao:2020ito,Hua:2020gnw,Ji:2020brr,Ji:2020jeb,LatticeParton:2020uhz,Lin:2020fsj,Lin:2020rxa,Lin:2020ssv,Shanahan:2020zxr,Shugert:2020tgq,Vladimirov:2020ofp,Zhang:2020dkn,Zhang:2020gaj,Zhang:2020rsx,Alexandrou:2021bbo,Alexandrou:2021oih,Bhattacharya:2021moj,Bhattacharya:2021rua,Constantinou:2021nbn,Dodson:2021rdq,Gao:2021hxl,Gao:2021dbh,LatticePartonLPC:2021gpi,Li:2021wvl,Lin:2021brq,Lin:2021ukf,Scapellato:2021uke,Schlemmer:2021aij,Shanahan:2021tst,Bhattacharya:2022aob,Constantinou:2022fqt,Deng:2022gzi,Ebert:2022fmh,Gao:2022iex,Gao:2022uhg,LatticeParton:2022xsd,LatticeParton:2022zqc,LatticePartonCollaborationLPC:2022myp,LatticePartonLPC:2022eev,Scapellato:2022mai,Schindler:2022eva,Zhang:2022xuw,Alexandrou:2023ucc,Avkhadiev:2023poz,Bhattacharya:2023jsc,Bhattacharya:2023nmv,Bhattacharya:2023tik,Cichy:2023dgk,Deng:2023csv,Gao:2023ktu,Gao:2023lny,Holligan:2023jqh,Holligan:2023rex,Ji:2023pba,LatticeParton:2023xdl,LatticePartonLPC:2023pdv,Lin:2023gxz,Zhao:2023ptv,Liu:2023onm,Avkhadiev:2024mgd,Baker:2024zcd,Bollweg:2024zet,Chen:2024rgi,Cloet:2024vbv,Ding:2024saz,Gao:2024fbh,Good:2024iur,Han:2024min,Holligan:2024umc,Holligan:2024wpv,Ji:2024hit,LatticeParton:2024mxp,LatticeParton:2024vck,LatticeParton:2024zko,Miller:2024yfw,Mukherjee:2024xie,Spanoudes:2024kpb,Zhang:2024omt,Bollweg:2025ecn,Bollweg:2025iol,Han:2025odf,Holligan:2025baj,LPC:2025spt,Zhang:2025hvf,Ji:2025mvk,Ji:2015jwa,Chen:2016fxx,Zhang:2018ggy,Braun:2018brg,Wang:2019tgg,Ji:2021uvr,Su:2022fiu,Zhu:2022bja,Yao:2022vtp,Pang:2024sdl,Han:2024cht,LatticePartonCollaborationLPC:2025vhd,Tan:2025ofx,Gao:2026hix,Gao:2026wlz,LPC:2026vyv,Zhang:2026lle,Grebe:2026qmt,LPC:2026mvw,NieMiera:2025inn,NieMiera:2025mwj,NieMiera:2025vcx,ChenChen:2025amm}, establishing it as a mature first-principles framework for lattice parton physics.

The ingredients of a precision LaMET analysis have also come under increasingly quantitative control~\cite{Zhao:2025oto}. The nonlocal Wilson-line operators are multiplicatively renormalizable~\cite{Ji:2017oey,Ishikawa:2017iym,Green:2017xeu}, with nonperturbative renormalization available in regularization-independent momentum-subtraction (RI/MOM)~\cite{Constantinou:2017sej,Stewart:2017tvs,Alexandrou:2017huk,Chen:2017mzz}, ratio-type~\cite{Orginos:2017kos,Braun:2018brg,Li:2020xml,Fan:2020nzz}, and hybrid~\cite{Ji:2020brr} schemes. Self-renormalization~\cite{LatticePartonLPC:2021gpi} is a practical strategy for determining the ultraviolet subtraction, applicable to the hybrid scheme as well as to ratio-type prescriptions. Leading-renormalon resummation (LRR) removes the associated ambiguity and linear power correction~\cite{Holligan:2023rex,Zhang:2023bxs}, while higher-order matching, renormalization-group resummation (RGR) and threshold resummation control perturbative corrections as well as large logarithms~\cite{Gao:2021hxl,Li:2020xml,Chen:2020ody,Su:2022fiu,Ji:2023pba,Cheng:2024wyu,Ji:2024hit,Ji:2025mvk}. Residual power corrections can be quantified through infinite-momentum extrapolation, yielding a systematically improvable moderate-$x$ window.  These precision-control techniques make LaMET mature enough to encode as an executable workflow, but their complexity also raises the barrier that such an agent is intended to lower. 

The operator definition of a quasi-distribution is not unique: operators that project onto the same light-front physics form a universality class~\cite{Ji:2020ect,Hatta:2013gta}. More recently, calculations based on a broader set of operator constructions within this universality class have been realized~\cite{Gao:2023lny,Zhao:2023ptv,Zhang:2026lle,Grebe:2026qmt}. These constructions can differ in their ultraviolet structure at finite momentum and therefore in their renormalization and power corrections. The conventional gauge-invariant construction employs Wilson lines, which introduce linear divergences and lead to an exponentially worsening signal-to-noise ratio at large separations. Quasi-observables defined in Coulomb gauge avoid Wilson lines and their linear divergences while preserving three-dimensional rotational symmetry, thereby simplifying lattice renormalization and improving the signal-to-noise ratio, especially for TMD observables in the nonperturbative large-transverse-separation region~\cite{Gao:2023lny,Zhao:2023ptv,Zhao:2025oto}. Both gauge-invariant and Coulomb-gauge realizations appear in the analyses supported by the present framework.

Turning lattice correlators into physical parton distributions is a multi-stage workflow. Once a theoretical scheme is specified, operations such as a Fourier transform with asymptotic extrapolation or a selected matching kernel are deterministic. Other stages require dataset-dependent judgment: selecting correlator fit windows and numbers of excited states, combining competing strategies through Bayesian model averaging, and choosing asymptotic, continuum, or infinite-momentum extrapolation ans\"atze. Experienced researchers traditionally make these decisions by iteratively inspecting fit diagnostics and intermediate results under physical constraints. The resulting reliance on expert intervention and repeated implementation work limits efficiency, reproducibility, and the broader adoption of the methodology.

Large language models (LLMs) with tool calling and multi-step reasoning capabilities~\cite{Yao:2022react} make it possible to encode this expert judgment as constrained, inspectable decisions. For researchers new to LaMET, an agent can expose a standard analysis pipeline, make intermediate physics visible, and support learning the methodology through use. For experienced practitioners, it can remove repetitive coding and routine strategy selection while preserving user intervention for consequential choices. More generally, packaging analysis knowledge together with a mature theoretical framework can lower the cost of adoption and aid in its dissemination.

The longer-term value of such a framework depends on what it contributes beyond the capabilities of the underlying model. As foundation models improve~\cite{Kaplan:2020scaling,Wei:2022emergent}, capabilities such as multi-step reasoning, tool use, and code execution are increasingly incorporated into general-purpose systems~\cite{Wei:2022chain,Schick:2023toolformer,DeepSeekAI:2025r1}. Agent projects built primarily around these generic capabilities therefore offer diminishing distinct value as the underlying models become more capable~\cite{Jimenez:2023swebench,Yang:2024sweagent}. Agents targeted at a narrow domain are less exposed to this form of obsolescence~\cite{Rein:2023gpqa}. The analysis practice of a field such as lattice parton physics resides largely in collaboration-internal code, and choices reported only in
summary form, and is therefore sparsely represented in the public corpora on which
general models are trained. The limiting factor is then not the model's reasoning
ability but its access to codified domain
practice~\cite{Boiko:2023coscientist,Bran:2024chemcrow,Lu:2024aiscientist}. A framework that supplies validated numerical implementations, explicit scheme and
convention constraints, and heuristics distilled from expert usage thus contributes what a stronger general model does not provide by itself. Recent work has also begun to apply LLM agents directly to lattice QCD, including agentic generation of measurement workflows~\cite{Gao:2026cpd} and an autonomous LaMET analysis of the Collins--Soper kernel~\cite{Tan:2026ier}. Parallel developments in high-energy phenomenology, such as agentic electroweak phase-transition analysis~\cite{Wang:2026jjn}, likewise encode a convention-sensitive multi-stage pipeline as an inspectable workflow. These studies show that domain-specialized agents are becoming practical in this field.

For a method-specific scientific agent to deliver these benefits in practice, its design should follow three principles. First, physical correctness requires a careful balance between LLM agency and deterministic numerical tools, together with end-to-end validation through reproductions of established analyses. Second, the pipeline should remain transparent, with inputs, outputs, and intermediate artifacts carrying clear physical meaning and supporting independent inspection. Third, the implementation should be structurally clear, accessible, and extensible, allowing established technical components to be reused while preserving user control over choices that require scientific judgment.

We present \lamet, an open-source Python framework designed around these conditions. Its central contribution is to formalize LaMET analysis knowledge as an auditable workflow that separates data-dependent decisions from deterministic computation. A complete analysis is declared as a directed acyclic graph (DAG) of stage jobs in a single manifest. Each stage runs a deterministic program workflow over typed numerical routines and delegates the data-dependent choices to the LLM through bounded \texttt{ask} operations, whose structured responses are validated against the stage's requirements before use. All numerical evaluations are performed by deterministic tools, and intermediate results are serialized as self-describing NetCDF artifacts with Markdown reports so that provenance remains inspectable independently of the agent.

The present release focuses on quark collinear distributions, including PDFs and DAs, and covers correlator analysis, renormalization, Fourier transformation, perturbative matching, continuum, physical pion mass and infinite-momentum extrapolations, and automated result review. We demonstrate the framework through four end-to-end LaMET analyses: pion PDFs in both the gauge-invariant and Coulomb-gauge formalisms, and pion and kaon DAs. These examples provide systematic reproductions of established analyses and validate the framework across distinct observables and formulations. Extensions to higher-dimensional distributions such as TMDs and GPDs, gluonic distributions, together with broader precision-control support, including threshold resummation, are planned for subsequent releases.

\section{LaMET Analysis Workflow} 
\label{sec:pipeline} 

\begin{figure*}[t]
    \centering
    \includegraphics[width=0.98\textwidth]{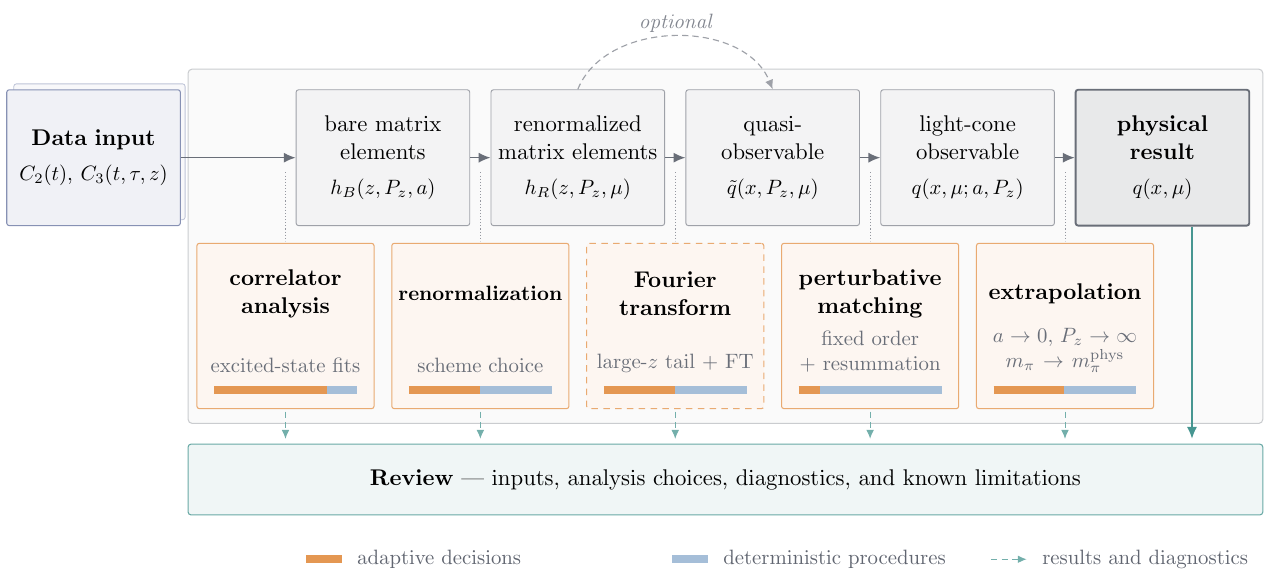}
    \caption{
    \justifying
    Scientific workflow of a generic LaMET analysis. The upper row shows the intermediate objects, the lower row the stage that maps one into the next. The bar in each stage box indicates schematically the split between data-dependent decisions requiring expert judgment (orange) and deterministic operations (blue). The Fourier transform is bypassed when the quasi-observable is already defined in momentum space; the review below the frame collects the record of every stage without altering the results.
    }
    \label{fig:lamet_workflow}
\end{figure*}

A LaMET analysis proceeds through a sequence of conceptually distinct stages that transform Euclidean lattice correlators into light-cone parton observables. \autoref{fig:lamet_workflow} summarizes this workflow, from correlator analysis and renormalization through Fourier transformation, perturbative matching, and extrapolation to the final physical results. It also highlights the intermediate numerical artifacts passed between stages and, importantly for the agent framework developed in this work, distinguishes steps that require substantial expert judgment from those that can be carried out by deterministic numerical tools.

The workflow shown in \autoref{fig:lamet_workflow} applies directly to the PDF and DA examples considered in this work and captures the common structure of more general LaMET analyses, including GPDs and TMDs, with additional kinematic variables and operator-specific systematics. From the perspective of an agent framework, the essential feature is that these stages differ not only in their numerical operations, but also in the type and degree of scientific judgment they require.

\paragraph{Correlator analysis:}
The correlator analysis is one of the most judgment-intensive stages of the LaMET workflow. It converts Euclidean correlation functions into the bare equal-time matrix element associated with the quasi-observable of interest. For example, in the quasi-PDF case, the relevant hadron matrix element is
\begin{align}
h_B(z,P_z,a)
=\langle H(P_z)|{\cal O}_{\Gamma}(z)|H(P_z)\rangle_B ,
\label{eq:workflow_bare_matrix}
\end{align}
where ${\cal O}_{\Gamma}(z)$ denotes the corresponding nonlocal operator.
Extracting the ground-state matrix element requires combining information from two- and three-point correlation functions, with several choices for how the analysis is organized. The matrix element can be constrained using the standard ratio of three- to two-point functions, a Feynman--Hellmann construction~\cite{Bouchard:2016heu}, or a combined analysis of both observables. The two-point spectrum can be determined in the same joint fit, propagated from a separate two-point fit through a chained analysis, or fixed to previously determined values.

More recently, Lanczos-based approaches~\cite{Wagman:2024rid,Hackett:2024xnx,Hackett:2024nbe,Abbott:2025yhm} have provided an alternative to conventional multi-state fitting: they construct a finite-dimensional representation of the transfer matrix in a Krylov subspace generated from the correlators and extract the relevant spectral information through eigenvalue decomposition. This reformulation can reduce the dependence on explicit choices of the number of states and nonlinear fit windows, although truncation of the Krylov space and statistical stability introduce their own analysis choices.

For a standard multi-state fit based on two- and three-point functions, the extraction follows from their spectral decompositions,
\begin{align}
\begin{aligned}
  C_2(t)
  &=
  \sum_n |A_n|^2 e^{-E_n t}+\cdots ,
  \\
  C_3(t,\tau,z)
  &=
  \sum_{m,n}
  A_m A_n^\ast
  \langle m|{\cal O}_{\Gamma}(z)|n\rangle
  e^{-E_m(t-\tau)}e^{-E_n\tau}
  \\
  &\quad+\cdots ,
\end{aligned}
  \label{eq:workflow_correlator_fit}
\end{align}
where conventional kinematic normalization factors are suppressed. The desired bare matrix element is isolated from the ground-state contribution, $m=n=0$, after the corresponding overlap and kinematic factors are removed.

In practice, this strategy has to be fixed together with the Euclidean-time windows entering each fit, the number of excited states retained, and whether competing fit models are combined through model averaging. None of these has a universal default, since what is appropriate depends on how the data behave, and every dataset has to be examined on its own terms. Nor are they settled in a single pass: each is proposed, tested against fit-quality and stability diagnostics, and revised in light of the outcome. This cycle of proposing, inspecting, and adjusting is well suited to an agent.

\paragraph{Renormalization:}
Renormalization removes the ultraviolet regulator dependence of the bare matrix element and defines the corresponding renormalized matrix element in a specified scheme.  For a multiplicatively renormalizable nonlocal operator, this relation can be written schematically as
\begin{align}
  h_R(z,P_z,\mu)
  =
  Z_{\cal O}^{-1}(z,\mu,a)\,
  h_B(z,P_z,a).
  \label{eq:workflow_renormalization}
\end{align}
Different renormalization prescriptions differ primarily in how the renormalization factor, or equivalently the ultraviolet subtraction, is determined. Common choices in LaMET calculations include RI/MOM-type schemes, in which the renormalization condition is imposed on off-shell partonic matrix elements, and ratio-type schemes, in which the ultraviolet factors are canceled using a hadronic reference matrix element.

In the ratio-type scheme, a commonly used reference is the corresponding zero-momentum matrix element.  Before imposing an overall normalization, the renormalized correlator is defined as
\begin{align}
  \widehat h_R^{\rm ratio}(z,P_z)
  =
  \frac{h_B(z,P_z)}{h_B(z,0)} .
  \label{eq:workflow_ratio}
\end{align}
Since the numerator and denominator contain the same nonlocal operator at the same separation $z$, their common $z$-dependent ultraviolet factors cancel.
What remains is an overall normalization, fixed by the convention appropriate for the observable.  Requiring the local matrix element to retain its normalization, $h_R^{\rm ratio}(0,P_z)=1$, for instance gives
\begin{align}
  h_R^{\rm ratio}(z,P_z)
  =
  \frac{h_B(0,0)}{h_B(0,P_z)}
  \widehat h_R^{\rm ratio}(z,P_z).
  \label{eq:workflow_ratio_normalized}
\end{align}

A limitation of the ratio-type prescription is that the zero-momentum matrix element in the denominator becomes increasingly affected by infrared effects at large distances.  Hybrid renormalization addresses this problem by retaining the ratio prescription at short distances while replacing the zero-momentum denominator at large distances with an explicit subtraction of the Wilson-line power divergence.  A representative unnormalized hybrid correlator can be written as 
\begin{align}
  \widehat h_R^{\rm hyb}(z,P_z)
  =
  \begin{cases}
  \displaystyle
  \frac{h_B(z,P_z)}{h_B(z,0)},
  & |z|\le z_s,
  \\[9pt]
  \displaystyle
  \frac{h_B(z,P_z)}{h_B(z_s,0)}
  e^{(\delta m+m_0)(|z|-z_s)},
  & |z|>z_s ,
  \end{cases}
  \label{eq:workflow_hybrid_ratio}
\end{align}
where $z_s$ is the switching separation.  The same overall normalization used for the short-distance ratio can also be applied to the complete hybrid correlator.

Implementing the long-distance branch of
Eq.~\eqref{eq:workflow_hybrid_ratio} requires the subtraction parameters $\delta m$ and $m_0$.  The coefficient $\delta m$ removes the Wilson-line linear divergence, while $m_0$ represents a residual scheme-dependent mass contribution when such a term is included in the chosen prescription.  Several strategies are used to determine these quantities.  In self-renormalization approaches, the dependence of the bare matrix elements on the lattice spacing and spatial separation is analyzed simultaneously to separate the power divergence from logarithmic and finite contributions.  In other implementations, the relevant subtraction parameters are obtained from fits to the coordinate-space matrix elements using a prescribed functional form over a selected range of $z$. The result therefore depends not only on the renormalization scheme itself, but also on the fit model, fit window, treatment of residual short-distance contributions, and, in the hybrid case, the choice of $z_s$.

Once the scheme, normalization convention, and subtraction parameters have been specified, the renormalization map is deterministic and can be applied sample by sample.  The scientific judgment lies in selecting among RI/MOM, ratio, hybrid, or related prescriptions; defining the reference and normalization conditions; determining the power-divergence subtraction; and assessing the stability of the associated fit and switching choices. These inputs should therefore be explicitly recorded and validated before the subsequent stages of the analysis.

\paragraph{Fourier transformation:} 

Lattice calculations provide the renormalized coordinate-space correlator only at a finite set of separations, and therefore over a finite range of the Ioffe time $\lambda=zP_z$.  Directly truncating the Fourier integral at the largest available separation amounts to imposing a sharp window in coordinate space.  If the correlator has not sufficiently decayed near this boundary, the resulting quasi-distribution can exhibit sizeable oscillation artifacts.  To control these finite-range effects, the coordinate-space data are first completed at large $|z|$ using the asymptotic long-distance expansion recently developed in Ref.~\cite{Ji:2026vir}, in which confinement leads to an exponentially decaying correlator multiplied by observable-dependent oscillatory and inverse-distance terms.

For example, for PDF- and GPD-like observables, a representative tail combining the leading-asymptotic (LA) and next-to-leading-asymptotic (NLA) orders can be written schematically as~\cite{Ji:2026vir}
\begin{align}
\begin{aligned}
h_R^{\rm tail}(z)
&\simeq
e^{-\Lambda |z|}
\bigg[
\sum_j A_j
e^{i(\phi_j\,{\rm sign}(z)+\omega_j z)}
\\
&\quad+
\frac{1}{|z|}
\sum_j A'_j
e^{i(\phi'_j\,{\rm sign}(z)+\omega'_j z)}
\bigg],
\qquad |z|\to\infty ,
\end{aligned}
\label{eq:workflow_tail}
\end{align}
where the first sum is the LA contribution, while the $1/|z|$ terms enter at NLA order.  Equation~\eqref{eq:workflow_tail} is illustrative but not universal: the number of terms and the allowed amplitudes, phases, and frequencies depend on the observable, external states, hadron, and flavor sector.  In particular, PDF and GPD correlators involve different momentum-dependent frequencies, while meson DA observables require their corresponding quasi-DA asymptotic forms. For Coulomb-gauge correlators, the large-distance behavior is not yet known explicitly; we supplement the tail ansatz with a power-law factor, $(z_0/|z|)^n$, following Ref.~\cite{Gao:2026hix}.

After completing the correlator at large $|z|$ and reconstructing the negative-$z$ region using the appropriate symmetry relation, the PDF-like quasi-distribution is obtained from
\begin{align}
  \tilde q(x,P_z,\mu)
  =
  P_z\int_{-\infty}^{\infty}\frac{dz}{2\pi}\,
  e^{ixP_z z} h_R(z,P_z,\mu).
  \label{eq:workflow_fourier}
\end{align}
Other channels instead use the corresponding sine, cosine, or symmetry-projected transforms. Once the phase convention and $x$ grid are fixed, no additional modeling enters the transform itself. Its stability can be tested by varying the tail-fit window, the asymptotic truncation, and the point at which the fitted tail replaces the lattice data, while monitoring residual oscillations in the resulting quasi-distribution.

Alternatively, for Coulomb-gauge correlators, the discrete Fourier sum can be incorporated directly into the lattice operator, so that the momentum-space quasi-distribution is extracted without first reconstructing the coordinate-space matrix element or modeling its large-$|z|$ tail~\cite{Grebe:2026qmt}. In this case, the separate tail-completion and Fourier-transform stages are bypassed, and the resulting quasi-distribution is passed directly to the matching stage.

\paragraph{Perturbative matching:} Once the quasi-distribution has been constructed in a renormalization scheme $\mathrm R$, it is related to the light-cone distribution through LaMET matching.  For a quasi-PDF, the fixed-order factorization formula can be written as
\begin{align}
  q(x,\mu)
  &=
  \int_{-\infty}^{\infty}\frac{\dd y}{|y|}\,
  C_{\mathrm R} 
  \left(
    \frac{x}{y},
    \frac{\mu_R}{\mu},
    \frac{yP_z}{\mu}
  \right)
  \tilde q\,(y,P_z,\mu_R)
  \nonumber\\
  &\quad
  +{\cal O}\!\left(
    \frac{\Lambda_{\rm QCD}^2}{P_z^2},
    \frac{M_H^2}{P_z^2}
  \right).
  \label{eq:workflow_matching}
\end{align}
Here $\mathrm R$ denotes the renormalization prescription used for the quasi-observable, such as an $\overline{\rm MS}$, ratio, RI/MOM, or hybrid scheme. The kernels used in this work are evaluated at next-to-leading order (NLO)~\cite{Yao:2022vtp,Baker:2024zcd,Chen:2024jkb,Gao:2026hix,Gao:2026wlz}.

The perturbative kernel can be improved beyond fixed order.  RGR evaluates the coefficient near its natural partonic scale, of order $2|y|P_z$, and evolves the result to $\mu$, resumming logarithms associated with the separation of these scales~\cite{Su:2022fiu}.
For Wilson-line observables, LRR resums the leading infrared renormalon and cancels the corresponding ambiguity against that in the Wilson-line mass subtraction. A consistent treatment therefore uses the same renormalon
prescription in both renormalization and matching~\cite{Zhang:2023bxs}.  Near the endpoint, threshold factorization separates the active-parton and spectator scales and resums the resulting threshold logarithms~\cite{Ji:2023pba,Ji:2024hit}.

These improvements leave the convolution structure unchanged and amount to replacing the scheme-specific fixed-order coefficient by the corresponding improved one,
\begin{align}
C_{\mathrm R}
\longrightarrow
C_{\mathrm R}^{\rm imp},
\end{align}
where $C_{\mathrm R}^{\rm imp}$ can combine fixed-order corrections with RGR, LRR, and threshold resummation.  The light-cone distribution is then reconstructed numerically from the quasi-distribution using this matching relation.  Its perturbative uncertainty is assessed by varying the matching order, resummation accuracy, and relevant scales.  The matched result is regarded as perturbatively reliable only when both $xP_z$ and $(1-x)P_z$ remain well above $\Lambda_{\rm QCD}$; the endpoint regions near $x=0$ and $x=1$ are therefore not included in the quoted range.

\paragraph{Extrapolation:}
The extrapolation stage combines the matched distributions obtained at finite lattice spacing, pion mass, and hadron momentum into the final physical result. A representative simultaneous ansatz is
\begin{align}
\begin{aligned}
  &q(x,\mu;a,P_z,m_\pi)
  \\
  &=\,
  q(x,\mu)
  +c_a(x)a^2
  +c_P(x)\frac{1}{P_z^2}
  +c_{aP}(x)(aP_z)^2
  \\
  &\quad
  +c_\pi(x)
   \bigl(m_\pi^2-m_{\pi,\mathrm{phys}}^2\bigr)
  +\cdots .
\end{aligned}
  \label{eq:workflow_extrapolation}
\end{align}
The leading powers in this expression follow from different expansions. For an $\mathcal O(a)$-improved lattice action and operator, the momentum-independent discretization error begins at $\mathcal O(a^2)$. The leading momentum-dependent discretization effect is correspondingly expected to scale as $(aP_z)^2$. If the action or the nonlocal operator is not fully $\mathcal O(a)$ improved, a term linear in $a$ can instead be allowed and
should be tested in the fit. For the standard leading-twist quasi-distributions considered here, the LaMET expansion gives leading finite-momentum corrections of order $\Lambda_{\rm QCD}^2/P_z^2$ and $m_h^2/P_z^2$. No independent $\mathcal O(1/P_z)$ contribution is expected for these operators. We therefore use $1/P_z^2$ as the leading finite-momentum dependence, while alternative power corrections can be tested when required by the observable or the data. The pion-mass dependence is written here in its simplest analytic form, while chiral-logarithmic terms or other parametrizations can be used when the available ensembles can constrain them. For each selected fit form, the extrapolation is repeated on every bootstrap or jackknife sample.

Statistical uncertainties follow from the resulting sample distribution of the physical-point distribution $q(x,\mu)$. Because the preceding stages are applied sample by sample, this error already includes the correlations generated by the correlator extraction, renormalization, Fourier transform, and matching.
Systematic uncertainties are estimated separately, by repeating the analysis under alternative, physically motivated choices and comparing those outcomes with the central determination.
The relevant variations are those already identified above: the Euclidean-time windows and spectral models used to extract the matrix element; the renormalization prescription and, for a hybrid scheme, the switching distance $z_s$; the large-$|z|$ tail model and fitting window of the Fourier transform; the matching order, resummation accuracy, and factorization scale; and the functional form of Eq.~\eqref{eq:workflow_extrapolation}, including mixed $(aP_z)^2$ or higher-power corrections.
When several alternatives probe the same source, their pointwise envelope is assigned to that component; independent components are then combined in quadrature and reported together with the statistical error.
The quoted result is this physical-point distribution together with the corresponding uncertainty budget, restricted to the interval of $x$ in which the matching remains perturbatively controlled.

The numerical analysis ends at this stage.  The final review collects the inputs, fit diagnostics, tables, figures, and analysis choices in one place.  It records missing inputs, unstable fits, inconsistent conventions, and known limitations of the lattice data, but does not alter the reported results.

Equations~\eqref{eq:workflow_bare_matrix}--\eqref{eq:workflow_extrapolation} also define the interfaces between stages of the LaMET agent.  Correlators and distributions are passed as resampled numerical data together with their metadata, and all numerical operations are performed by deterministic tools.  The language model is consulted where a stage requires a data-dependent choice among the supported analysis options, while the consistency of schemes and conventions and the identification of missing inputs are handled by explicit program checks.  Since an analysis can begin from raw correlators or from an intermediate result, the stages are implemented as independent jobs.  This modular structure is described next.

\section{Execution Model}
\label{sec:execution}

\begin{figure*}[tbp]
\centering
\includegraphics[width=0.93\textwidth]{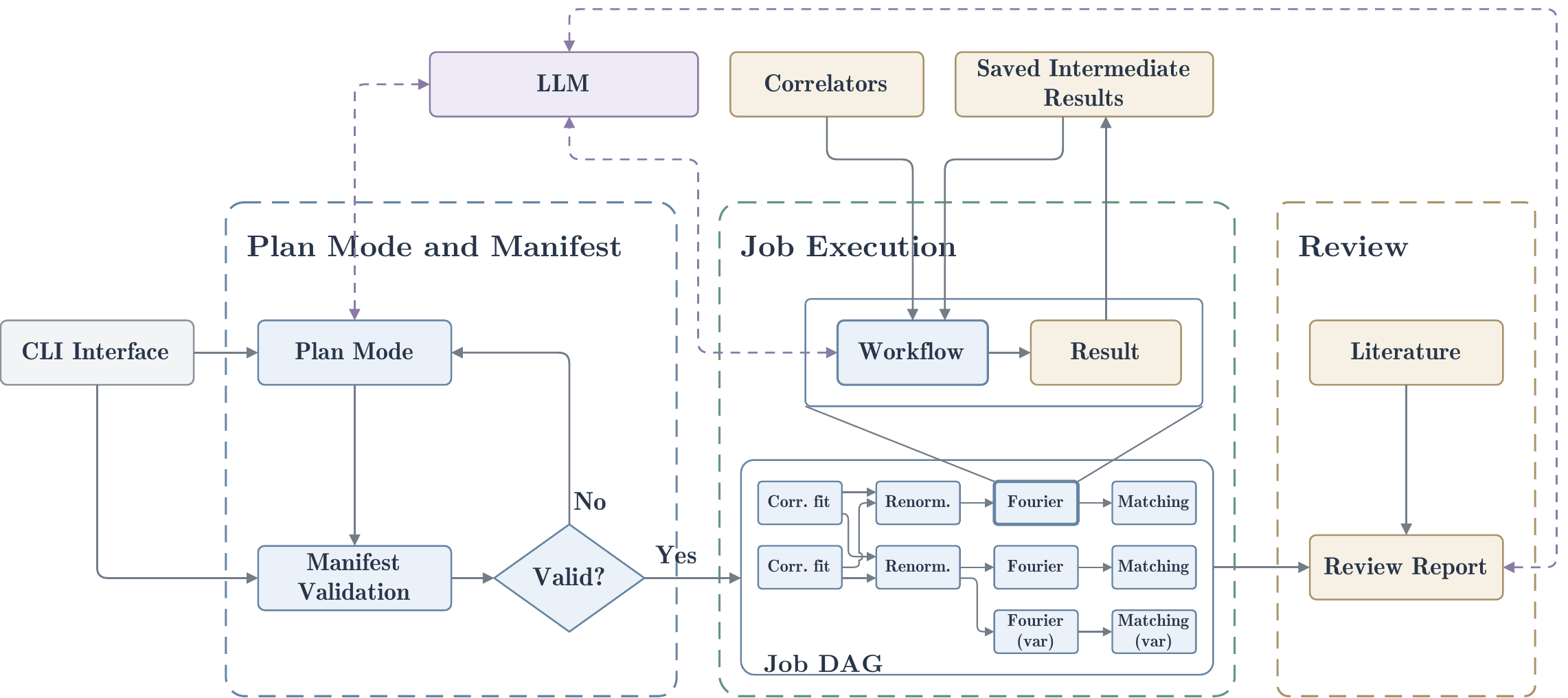}
\caption{
\justifying
Execution model of \lamet. Plan Mode turns the user's physics objective into a manifest, which is executed only once it satisfies the contract; validation
errors return to Plan Mode. The validated manifest is expanded into a job DAG
whose jobs the agent then executes in a single ordered pass consistent with that dependency order under deterministic numerical tools. One
job (heavier outline) is expanded above as an example of how any job is
organized: it loads external correlators or the results of preceding jobs, runs the stage workflow, and returns a result that is saved for dependent jobs and for later runs. The crossed dependencies in the first two columns indicate that a job may consume several upstream results; jobs labeled \emph{var} are those generated for systematic variations. Dashed lines indicate the three points at which the LLM is used: the initial planning dialogue, \texttt{ask} operations within a stage workflow, and the final Review. The Review compares the recorded results with the literature, writing its recommendations without changing the manifest or any numerical artifact. 
}
\label{fig:lamet-analysis-workflow}
\end{figure*}

\lamet\ organizes a LaMET analysis into preparation, numerical job
execution, and interpretation. Through dialogue with the user, the LLM helps
translate physics objectives into concrete analysis choices recorded in a
manifest. The agent checks this specification, carries out the calculations in
their dependency order, and, when requested, reviews the results in the context
of the relevant literature. Physical validation, propagation of systematic
variations, and recorded numerical evidence support reliable numerical results
and allow the user to trace how analysis choices affect the final result.
\autoref{fig:lamet-analysis-workflow} shows the execution model of the agent.

\subsection{Plan Mode and Manifest}

\paragraph{Plan Mode} Plan Mode lets the user discuss and revise an existing
analysis with an LLM. A typical session starts from a manifest and its
accompanying correlator description file, since generating both from scratch
requires substantial information about the calculation and input data, which
is not easy to prepare. The user can ask questions about the current workflow
or describe changes to the observable, numerical method, or treatment of
uncertainties in scientific terms. The LLM answers the query and, when the user
requests a change, translates the natural language into manifest edits.

\begin{figure*}[tbp]
\centering
\includegraphics[width=0.96\textwidth]{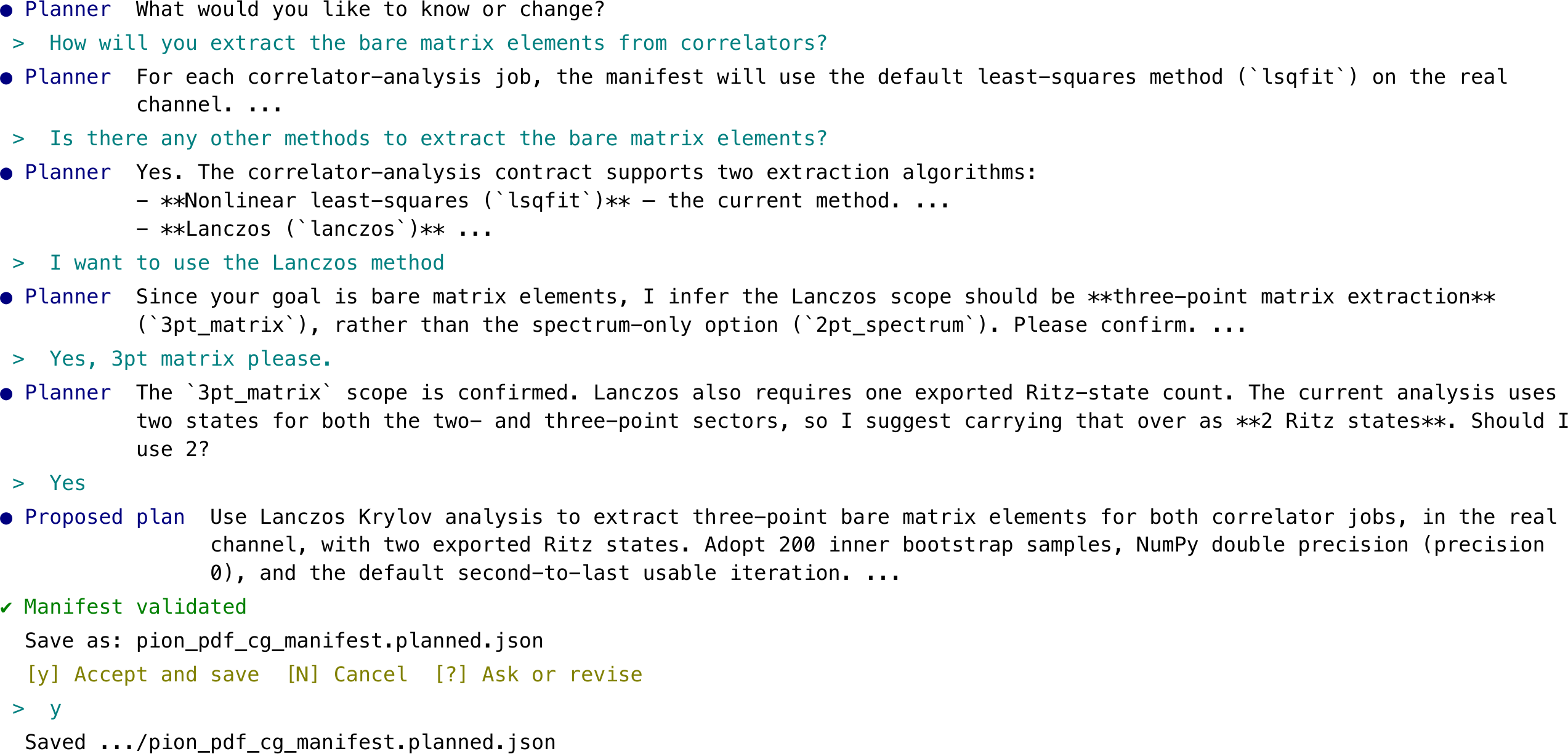}
\caption{
\justifying
A conversation between the planner and the user asking for a change in the matrix
elements extraction method from least-squares fitting to the Lanczos algorithm.
The user asks questions and describes changes in natural language, and the planner
answers the question or applies the revision to the manifest.
}
\label{fig:lamet-plan-conversation}
\end{figure*}

After each edit, validation feedback guides the next exchange, including any
questions about missing settings or requests for the user's confirmation to use
default values. The user can revise the plan over several turns and undo changes
before finally accepting it. Once the user accepts a valid configuration, Plan
saves the revised manifest. Plan mode is optional for a valid manifest, and will
be entered if validation errors are found.
\autoref{fig:lamet-plan-conversation} illustrates an example of a user asking for
a change in the correlator analysis stage through a conversation with the plan mode.

\paragraph{Manifest} The manifest specifies the target observable, input data,
physical and statistical conventions, calculation stages, and optional
variations for estimating systematic uncertainties. A stage represents a step in
the analysis, such as extracting matrix elements from correlators or matching a
quasi-distribution to a light-cone distribution. Individual jobs are instantiated
with specified inputs and parameter choices, and then apply the corresponding step.

Before any calculation begins, the manifest is checked against a set of
requirements called a contract. These requirements, along with some extra
stage-specific validators, describe supported options,
which parameters they need, and which physical conditions must hold.
For example, choosing least-squares fitting introduces requirements for fitting
windows, while choosing Lanczos analysis introduces a different set of
parameters. Validation applies available defaults and identifies invalid values or
incompatible setting combinations, including inconsistencies between input kinematics or
resampling choices.
Each validation rule includes a human-readable explanation of the physical or statistical reason for the corresponding requirement. When validation fails, this explanation appears with the error message and is provided to the LLM to help it formulate appropriate planning questions.

The manifest also supports systematic-error estimation by describing variations
of the central analysis. Before execution, the agent expands the central job with
these variations into additional jobs and validates the expanded configuration.
These dependency relations between these jobs form the DAG of the analysis.
The agent executes the jobs in a single pass through the authored
stage and job order, which validation requires to be consistent with that DAG:
every declared input must name an upstream job that appears earlier in the
manifest or an external artifact. The declared graph therefore fixes what each
job consumes, while the authored order fixes when it runs.

\subsection{Job Execution}

\paragraph{Workflow} A job begins when its required inputs are available.
It loads external data or receives results from previous jobs, checks that
they meet the stage's requirements, and prepares them for the selected numerical
method. Parameters specified in the manifest determine the analysis workflow,
including the calculation branch, the fitting model, the matching kernel, and so on.

The workflow then performs the calculation and examines its diagnostics. For a
fitting procedure, this includes assessing candidate fits and selecting the
result according to the stage's quality criteria. A job stops when its
numerical procedure fails outright.

In particular, the perturbative matching stage supports both built-in and
user-defined kernels. The agent determines the kernel from the input data's
parton type, observable, gauge construction, and operator, together with the
matching scheme, perturbative order, and resummation choices specified in the
manifest. Users can add a numerical implementation and its formula documentation
under the corresponding naming convention, making it available for that
combination of physical settings. Kernel parameters, such as the renormalization
scale, remain configurable through the manifest. Once the input data are loaded
and the kernel is identified, the agent checks the supplied parameters against
its requirements before performing the matching calculation.

\paragraph{\texttt{ask}} Some analysis choices, such as fitting ranges, are not
uniquely determined by the data and physical constraints. An analyst would often
inspect the signal, compare fitting strategies, and assess their stability before
choosing these parameters. An \texttt{ask} operation is defined as a Python function
to be directly launched by the agent which pushes prompts and data to an LLM,
and requests a structured output of recommended parameters from the LLM. It is
designed to delegate a specific parameter-selection task to the LLM to reduce the
need for manual inspection of data.

When the workflow reaches a step that requires a recommendation, it assembles
the numerical summaries, physical guidance, requested parameters, and
constraints on their values. Then the agent validates the structured response,
parses these parameters, and applies them. Users can supply these parameters in the
manifest or leave supported values for recommendation at runtime.

For example, in Fourier-tail fitting, the candidate windows may be enumerated
in the manifest or, when left open, recommended by the LLM subject to the
stage constraint that they begin at measured separations of at least
$0.5\,\mathrm{fm}$. When no candidate
meets the stage's fit-quality threshold, the workflow launches the \texttt{ask}
again for another set of candidate windows. If the threshold cannot be satisfied
after several attempts, it publishes the best available result ranked by fit quality
and records a warning with it, rather than disturbing the whole analysis. Such a degraded
result is always visible in the Fourier transform report.

\paragraph{Data structure} Jobs exchange numerical results through
\texttt{EnsembleData}, which wraps an
\texttt{xarray.DataArray}~\cite{Hoyer_xarray_N-D_labeled_2017} and carries
statistical samples and physical metadata.
\texttt{xarray.DataArray} uses labeled dimensions and coordinates to support a metadata-aware array and operations. We can name an array dimension with physical meaning, which helps the agent operate on physical names instead of indices. We can also name coordinates in physical units, and both the user and the agent can check the data or plot figures without having to convert the units.
Besides the feature from \texttt{xarray.DataArray}, \texttt{EnsembleData}'s common interface also handles binning,
jackknife or bootstrap resampling, and uncertainty estimation with the standard
errors appropriate to each representation. Keeping the samples available allows
subsequent calculations to retain correlations, while named coordinates and
ensemble information record which physical quantities the arrays represent.

The array's \texttt{attrs} records additional physical information, including
the hadron identifier, the inserted current's operator and parton content,
polarization, momentum, and renormalization conventions. Later stages use these
attributes to check input compatibility and select the appropriate physical
formulas. For example, the Fourier stage uses the hadron and observable to
distinguish the asymptotic tail families of the PDF- and DA-like channels.

\paragraph{Systematics} Systematic-error studies are declared in the manifest's \texttt{systematics} object, which expands central jobs C1 into additional variation
jobs A1 within the same manifest and the same run. Each A1 job inherits the C1's
inputs and parameters, and each variant modifies requested parameters at the stage
where the corresponding systematic effect enters, so that the resulting spread is
attributable to those analysis choices. A1 jobs also inherit C1's dependency,
thus downstream variation jobs A2 from the downstream central jobs C2
use the corresponding A1's outputs, so the effect of a changed analysis choice is
carried through the subsequent calculations under the same variation label.
For example, several shifts of a Fourier-tail fitting window generate several
Fourier jobs from one central job. Their outputs feed corresponding matching
and extrapolation jobs, producing a set of final distributions to compare with
the central result. All these jobs run through the usual stage workflows.

At extrapolation, the uncertainty budget groups results by source. A single
alternative contributes its absolute deviation from the central value, while
multiple alternatives contribute their pointwise maximum-minus-minimum range.
Contributions from different sources are combined in quadrature, then combined
with the statistical uncertainty. The budget records each contribution and the
jobs used to compute it, allowing the user to trace an uncertainty source back
to the corresponding analysis variation.

\paragraph{Results and Reuse} Stage reports collect the completed jobs' results,
plots, physical metadata, and diagnostics. For example, at the end of the correlator analysis stage, the user can
check the report for selected parameters alongside fit-quality measures and the fitted
bare matrix elements. The recorded LLM transcript shows which parameters were
recommended, whether they were revised, and which choices were used in the
reported calculation.

Every numerical procedure in a stage workflow is deterministic. The manifest
declares a single run-wide random seed, from which each job derives its own
independent random stream, so that repeating a run with the same manifest, the
same input data, and the same execution environment
reproduces the reported results. The only exception is a parameter left to an
\texttt{ask} operation: a new model request may return a different choice,
so such parameters must be written back into the manifest before a reproduction
run. The stage reports record the recommended values for exactly this purpose.

The results are checked against the input compatibility conditions carried
by the metadata and the numerical diagnostics of each stage. These checks
address different possible failures, such as inconsistent kinematics or units
and unsuccessful fits. Reports of all stages with the systematic-error study and the final Review support an assessment of
physical validity together.

Numerical stages save their \texttt{EnsembleData} results as intermediate output files
in NetCDF4 format through \texttt{xarray}. These files retain numerical values, named
dimensions, coordinates, and physical attributes, together with ensemble
information and the sampling representation. They provide a record of the
intermediate calculations for inspection and can serve as inputs to a later
run if the user references these results and specifies them as the input data.
For example, a study of matching choices can start from saved transformed distributions
without repeating the extraction from raw correlators.

\subsection{Review}

\paragraph{Literature} The reference catalog is prepared before Review. The
Literature module searches INSPIRE for LaMET-related papers and downloads the
corresponding arXiv articles. A locally hosted language model reads the whole paper,
and then assigns structured labels for observables,
parton content, polarization, kinematics, numerical methods, and lattice setups.

Review matches these labels to the physical metadata and analysis settings of
the current calculation to obtain candidate references. The prepared catalog
can be reused across analyses. For a selected paper, text is read from the
local collection or retrieved online. If full text is unavailable, the catalog
summary is supplied instead. Retrieval status and any text truncation are
recorded, making clear which material was available for the literature
comparison.

\paragraph{Review Report} Review brings together the numerical results, stage diagnostics,
and uncertainty estimates to discuss what the completed analysis establishes
about the target observable. Within the scope specified by the manifest, it
relates the final distribution to the choices made during extraction,
transformation, and matching. This gives the reader an account of how those
choices affect the interpretation of the result, including limitations visible
in the fit diagnostics and the systematic variations that were evaluated.

The agent first collects the selected results and stage reports and checks
consistency across the calculation chain. The report explains the consequences
of any incompatible physical metadata and identifies conclusions that cannot
be assessed with the available information. Statistical errors and the declared
systematic-error budget provide the basis for discussing the uncertainty of
the result. Sources not examined by the analysis remain limitations and will be
identified when relevant.

With the numerical summaries, consistency findings, and candidate references,
the LLM uses \texttt{tool} calls to inspect selected results and read papers.
Literature comparisons provide
methodological background and help interpret similarities or differences in
physical behavior, taking account of the observables, lattice setups, and
analysis methods involved. Numerical statements about the current calculation
are based on its recorded outputs, while the papers supply the context for
qualitative comparison. Based on available information, the report will judge
whether further calculations or systematic variations are needed.

\section{Case Study}
\label{sec:case_study}

To demonstrate both the completeness and versatility of \lamet\ in carrying out LaMET analyses from lattice correlation functions to physical distributions, we release four end-to-end examples spanning representative scenarios from published calculations. A manifest declares the job DAG along with the run metadata and per-stage parameters. Once validated, the agent executes the declared stages without further user intervention, consulting the LLM where a stage requires parameter recommendations, and produces the final distributions together with the intermediate NetCDF artifacts, stage reports, and the final review report, as described in \autoref{sec:execution}. 

The four examples are chosen to span the principal axes of the current collinear LaMET analyses. They cover both PDFs and DAs, both Coulomb-gauge (CG) and gauge-invariant (GI) operator constructions, both pion and kaon states, and both single-ensemble and multi-ensemble workflows. Concretely, the pion valence PDFs in the CG and GI formulations follow Ref.~\cite{Gao:2023lny}, whose PDF extraction rests on a single $a=0.06$~fm ensemble at fixed $P_z$ (a finer ensemble enters there only as a zero-momentum renormalization cross-check). The pion and kaon DAs~\cite{LatticeParton:2022zqc}, by contrast, combine three lattice spacings and three momenta into a joint continuum and infinite-momentum extrapolation. A variant of the CG PDF manifest additionally replaces the least-squares
correlator analysis with a nested-bootstrap Lanczos extraction of the three-point
matrix element.

The manifest parameters are not an exact reproduction of the published setups. In recent years, the LaMET framework has continued to evolve; a relevant example is the improved treatment of large-distance correlations through asymptotic analysis~\cite{Chen:2025cxr,Ji:2026vir}, which is used in the Fourier-transform stage of the present examples. Independently, several stages, most notably correlator analysis, do not admit a unique, universally preferred parameter choice: fit windows, the number of excited states, and related cuts remain data-dependent judgments. Therefore, the correlator analysis is not fixed to the reference settings in the following examples. That non-uniqueness is itself a reason to treat systematic error estimation as an integral part of the workflow. Consequently, the comparisons below are intended as reproductions that are consistent with the published results, but not identical to them at the level of every intermediate prescription.

\subsection{Pion Parton Distribution Functions}

The pion valence PDF is a standard benchmark for a collinear LaMET analysis, and Ref.~\cite{Gao:2023lny} determines it in two independent ways within the same lattice setup. In the LaMET framework, the light-cone distribution is accessed from an equal-time correlator of a boosted pion with $P^\mu=(P^t,0,0,P_z)$. The GI construction joins the quark fields by a straight Wilson line $W(z, 0)$, whereas the CG construction imposes $\vec\nabla\cdot\vec A=0$ instead,
\begin{align}
\begin{aligned}
 \tilde h_{\mathrm{GI}}(z,P_z)&=\frac{1}{2P^t}\langle\pi(P)|\bar\psi(z)\gamma^t W(z,0)\psi(0)|\pi(P)\rangle,
 \\
 \tilde h_{\mathrm{CG}}(z,P_z)&=\frac{1}{2P^t}\langle\pi(P)|\bar\psi(z)\gamma^t\psi(0)|_{\vec\nabla\cdot\vec A=0}|\pi(P)\rangle.
\end{aligned}
\label{eq:quasi_pdf_me}
\end{align}
Both approach the same light-cone limit under an infinite boost and therefore belong to the same universality class~\cite{Hatta:2013gta,Ji:2020ect}, but they differ in ultraviolet structure: $W(z,0)$ carries a linear divergence and the associated renormalon, which the CG operator does not~\cite{Gao:2023lny}. The remaining definitions and the lattice setup follow Ref.~\cite{Gao:2023lny}. For the present purpose, this contrast is what makes the example useful: the two analyses share one workflow but must take different renormalization and matching paths through it.

The two runs are defined by separate manifests for the CG and GI analyses. Each consumes its own set of two- and three-point correlators from Ref.~\cite{Gao:2023lny}. Both of them were built from the same quark propagators on the $48^3\times64$ highly improved staggered quark (HISQ) ensemble at $a=0.06~\mathrm{fm}$, and both runs propagate 109 jackknife samples through every stage. Each run pairs a rest-frame job with a boosted one: $n_z=5$ ($P_z=2.15~\mathrm{GeV}$) for CG and $n_z=4$ ($P_z=1.72~\mathrm{GeV}$) for GI. Neither manifest declares an extrapolation stage, so the comparison is made at finite $P_z$ on this single ensemble against the reference results at the corresponding momentum.

\begin{table*}[t]
\caption{
\justifying
Physical setup, resolved settings, and fit diagnostics for the Coulomb-gauge and gauge-invariant pion-PDF runs, as reconstructed from the resolved manifests and stage reports. Where a manifest declares several candidates, the entry below is the choice the agent returned; the GI two-point windows and Fourier interval are single authored values. The correlator-fit $\chi^2/\mathrm{dof}$ is quoted separately for the rest-frame and boosted jobs of each run. The CG $z$ candidates are spaced by the lattice spacing $a=0.06~\mathrm{fm}$. The setup is stated in physical terms and implementation identifiers are shown only for kernels.
}
\label{tab:pion_pdf_settings}
\centering
\footnotesize
\setlength{\tabcolsep}{3pt}
\renewcommand{\arraystretch}{1.18}
\begin{tabular}{@{}p{32mm}p{67mm}p{67mm}@{}}
\hline
 & Coulomb gauge & Gauge invariant \\
\hline
Momenta &
$n_z=0$ / $n_z=5$, $P_z=2.15~\mathrm{GeV}$
&
$n_z=0$ / $n_z=4$, $P_z=1.72~\mathrm{GeV}$
\\
\noalign{\smallskip}

Correlator fit &
joint 2pt$+$3pt ratio, 2 states
&
joint 2pt$+$3pt ratio, 2 states
\\

\quad fit window &
rest : $t\in[3,12)$; $t_{\mathrm{sep}}=\{8,10,12\}$; $\tau_{\mathrm{cut}}=3$ \newline
boosted : $t\in[3,12)$; $t_{\mathrm{sep}}=\{8,10,12\}$; $\tau_{\mathrm{cut}}=3$
&
rest: $t\in[5,12)$; $t_{\mathrm{sep}}=\{8,10\}$; $\tau_{\mathrm{cut}}=3$\newline
boosted: $t\in[6,12)$; $t_{\mathrm{sep}}=\{8,10\}$; $\tau_{\mathrm{cut}}=3$
\\

\quad $\chi^2/\mathrm{dof}$
&
$0.070$ / $0.311$
&
$0.671$ / $0.208$
\\
\noalign{\smallskip}
\hline
\noalign{\smallskip}

Renormalization
&
hybrid scheme with an external rest-frame denominator; $z_s=0.18~\mathrm{fm}$
&
hybrid scheme with an external rest-frame denominator; $z_s=0.18~\mathrm{fm}$
\\

\quad $\delta m$, $m_0$
&
$0$, $0$
&
$0.5207~\mathrm{GeV}$, $0.1232~\mathrm{GeV}$
\\
\noalign{\smallskip}
\hline
\noalign{\smallskip}

Asymptotic region &
$[0.72,1.44]~\mathrm{fm}$
&
$[0.66,1.26]~\mathrm{fm}$
\\

\quad tail model &
LA/NLA continuation with \texttt{linear};\newline
$\chi^2/\mathrm{dof}=0.003$
&
LA/NLA continuation with \texttt{linear};\newline
$\chi^2/\mathrm{dof}=0.051$
\\
\noalign{\smallskip}
\hline
\noalign{\smallskip}

Matching
&
NLO hybrid matching at $\mu=2~\mathrm{GeV}$
&
NLO hybrid matching with LRR at $\mu=2~\mathrm{GeV}$
\\
\noalign{\smallskip}

\hline
\end{tabular}
\end{table*}

What a manifest specifies is a candidate space: two-point fit windows, source-sink separations and insertion-time cuts, Fourier ranges $(z_{\min},z_{\max})$, LA and NLA tail models, and prior-width scales. Within each stage the agent evaluates the declared candidates on sample-average data and applies the fit-quality gate $Q\ge0.05$. In correlator analysis a candidate below the gate triggers a bounded round of window revision; the best available fit is ranked by $Q$ and published with a recorded warning if no candidate clears the gate. In the Fourier stage, the candidates failing the gate are discarded, and the survivors are ranked by their Bayesian evidence. 
The selected window, Fourier range, and tail model are then held fixed across all resamples. \autoref{tab:pion_pdf_settings} records the resolved physical setup together with those selections and their fit diagnostics: correlator windows and $\chi^2/\mathrm{dof}$, hybrid-renormalization parameters, the retained asymptotic interval, and the matching kernel.

The two runs do not converge on the same settings. Both adopt a joint two-state 2pt$+$3pt-ratio fit, but the Euclidean windows differ. The CG analysis keeps $t\in[3,12)$ and $t_{\mathrm{sep}}=\{8,10,12\}$ in the rest frame and in the boost, whereas the GI rest-frame job uses $t\in[5,12)$ and the boosted job uses $t\in[6,12)$, each with only $t_{\mathrm{sep}}=\{8,10\}$. All four windows pass the quality gate at every $z$, with $\chi^2/\mathrm{dof}=0.070$--$0.671$, and no production resample falls below the threshold. In the Fourier stage the CG job retains data out to $z_{\max}=1.44~\mathrm{fm}$, while the GI job stops at $1.26~\mathrm{fm}$, consistent with the better long-distance precision expected without a Wilson line; both select an LA tail with prior width $3$, giving $\chi^2/\mathrm{dof}=0.003$ (CG) and $0.051$ (GI). Hybrid renormalization is applied in both cases at $z_s=0.18~\mathrm{fm}$, with vanishing mass counterterms for CG and $(\delta m,m_0)=(0.5207,0.1232)~\mathrm{GeV}$ for GI. The matching kernels are then NLO hybrid and NLO hybrid with LRR, respectively. The review stage closes both runs with no consistency errors and a single benign warning, the expected change of $x$ grid between the Fourier and matching stages.

\begin{figure}[t]
  \centering
  \includegraphics[width=\linewidth]
    {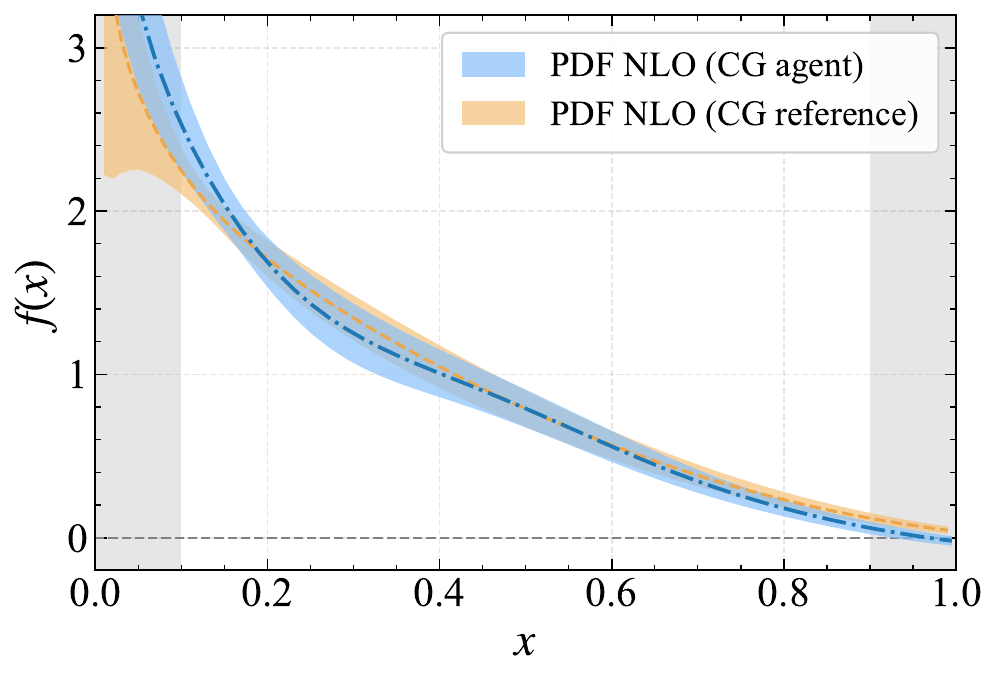}
  \caption{Coulomb-gauge pion light-cone PDFs from \lamet\ compared with the
  published results of Ref.~\cite{Gao:2023lny}, on the
  $a=0.06~\mathrm{fm}$ ensemble at $n_z=5$ ($P_z=2.15~\mathrm{GeV}$).
  The PDFs are matched at NLO. Shaded bands denote propagated statistical
  uncertainties. The gray regions indicate the estimated endpoint regions.} 
  \label{fig:pion_pdf_cg_stage_validation}
\end{figure}

\begin{figure}[t]
  \centering
  \includegraphics[width=\linewidth]    {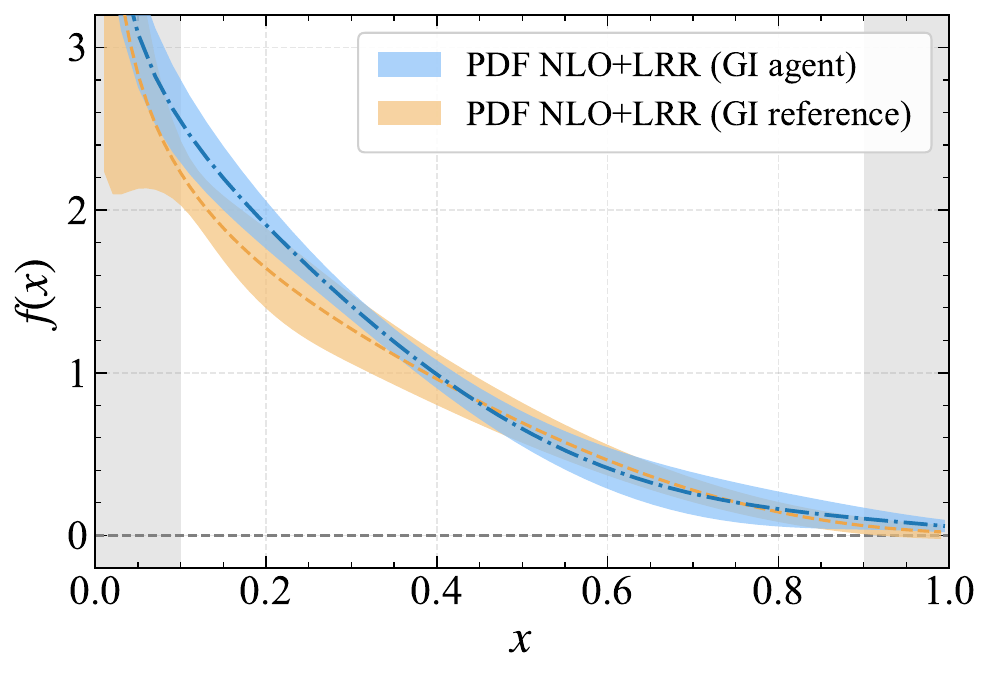}
  \caption{Gauge-invariant pion light-cone PDFs from \lamet\ compared with
  the published results of Ref.~\cite{Gao:2023lny}, on the same ensemble at
  $n_z=4$ ($P_z=1.72~\mathrm{GeV}$). The PDFs are matched at NLO with LRR.
  Shaded bands denote propagated statistical uncertainties. The gray regions indicate the estimated endpoint regions.}
  \label{fig:pion_pdf_gi_stage_validation}
\end{figure}

\autoref{fig:pion_pdf_cg_stage_validation} and
\autoref{fig:pion_pdf_gi_stage_validation} compare the resulting matched
light-cone valence PDFs with the reference results. On the GI side, the agent
applies the LRR hybrid kernel~\cite{Zhang:2023bxs}, consistent with the kernel
used in the published GI analysis shown in
\autoref{fig:pion_pdf_gi_stage_validation}. The two analyses nevertheless
differ in several implementation details. First, the agent uses
$z_s=0.18~\mathrm{fm}=3a$, whereas Ref.~\cite{Gao:2023lny} uses $4a$; the
$z_s=\{2,3,4\}a$ comparison reported in Ref.~\cite{Gao:2023lny} shows this
difference to be immaterial away from the discretization region. Second, the
Fourier stage models the unresolved tail with the LA and NLA asymptotic
forms~\cite{Chen:2025cxr,Ji:2026vir}, which differ from the exponential model
of the original analysis. Residual differences, most visibly in the GI PDF, should be attributed primarily to the different parameter settings in the ground state fits. At finite hadron momentum, the LaMET expansion suffers large power corrections in the endpoint regions, where either $xP_z$ or $(1-x)P_z$ approaches the nonperturbative scale $\Lambda_{\rm QCD}$. For the momenta used in these analyses, taking $\Lambda_{\rm QCD}=0.2~\mathrm{GeV}$ gives the rough estimate $\Lambda_{\rm QCD}/P_z\sim 0.1$. We therefore shade the regions $x<0.1$ and $x>0.9$ in gray throughout the following figures. The same regions are also shown for the $P_z\to\infty$ extrapolated results, since these results are obtained from finite-momentum inputs and remain less constrained near the endpoints. These boundaries should be understood as approximate guides rather than precise limits of the reliable region.

After the numerical stages have written their artifacts, an
LLM-supported review stage synthesizes the completed run while leaving every
result intact. As described in \autoref{sec:execution}, the reviewer receives the ordered job outputs and stage reports, a deterministic consistency audit along manifest
edges, and a bounded set of reference papers for methodological context. Every numerical claim is traced back to those artifacts, and the literature informs how they are interpreted.

In both pion-PDF examples the audit is clean apart from one warning: the
distribution grid changes from the symmetric Fourier interval $x\in[-2,2]$ to
the positive light-cone interval at matching, which the reports read as the
intended valence projection. Both reviews then report a valence-like matched
PDF that is positive and smoothly decreasing, and attribute the small
$\chi^2/\mathrm{dof}$ to the internal consistency between the selected models
and the data, reserving the word validation for an external test. The reviews also delimit the reach of these single-ensemble, single-boost
runs: a continuum or multi-momentum limit, a systematic-error budget, and a precision determination of the endpoints are identified as targets for a dedicated follow-up campaign, and the selected literature is kept in its methodological role.

\subsection{Meson Distribution Amplitudes}

Beyond pion PDFs, meson DAs characterize how the longitudinal momentum of a meson is shared among the quarks in its leading Fock component, providing another important class of collinear partonic observables. Ref.~\cite{LatticeParton:2022zqc} determines the pion and kaon DAs within the LaMET framework, where the DAs are accessed through equal-time gauge-invariant nonlocal correlators of boosted mesons,
\begin{align}
 \tilde h_M^{\Gamma}(z,P_z)&=\langle 0|
 \bar\psi_1(z)\Gamma W(z,0)\psi_2(0)|M(P)\rangle,
 \label{eq:quasi_da_me}
\end{align}
with $\Gamma=\gamma_z\gamma_5$ for the pseudoscalar channel. This correlator yields the meson DA $\phi_M(x)$ after renormalization, Fourier transformation, perturbative matching, and physical extrapolation. In the isospin limit, charge conjugation gives $\phi_\pi(x)=\phi_\pi(1-x)$, whereas the unequal light- and strange-quark masses lead to an asymmetric kaon DA. The detailed definitions and the lattice setup can be found in Ref.~\cite{LatticeParton:2022zqc}. As examples of \lamet, the pion and kaon DA cases exercise self-renormalization, multi-ensemble analysis, and the estimation of systematic uncertainties.

The pion and kaon analyses use the original gauge-invariant two-point correlators provided in Ref.~\cite{LatticeParton:2022zqc}. The available data consist of the nonlocal correlator together with its aligned $z=0$ slice, and the latter is used as the denominator in constructing the quasi-DA (qDA) ratio, following the original analysis. This differs from a normalization based on a conventional symmetric two-point function: the $z=0$ slice shares the operator construction and asymmetric source-sink smearing structure of the nonlocal correlator and therefore leads to a correspondingly different fit form for the ratio. Both the pion and kaon manifests include the $96^3\times192$, $64^3\times96$, and $48^3\times64$ HISQ ensembles at lattice spacings $a=0.0574$, $0.0882$, and $0.1213~\mathrm{fm}$, respectively, with momentum indices $n_z=6,8,10$. Across the three spatial volumes, these correspond to momenta $P_z=1.28\text{--}1.35$, $1.70\text{--}1.80$, and $2.13\text{--}2.25~\mathrm{GeV}$, respectively. Each analysis propagates 100 bootstrap samples through the full pipeline. 

At the correlator level, the pion example uses one-state fits to the real and imaginary components of the qDA ratio, whereas the kaon example performs a simultaneous fit to the nonlocal qDA correlator and its corresponding $z=0$ two-point correlator, a choice made to improve fit stability. Renormalization follows the self-renormalization procedure of Ref.~\cite{LatticeParton:2022zqc}: the renormalization factor is determined from the zero-momentum pion quasi-PDF matrix element, with the relevant coefficients replaced by their meson-DA counterparts following the same prescription. The subsequent stages include the choice of coordinate-space fitting range, asymptotic analysis of the large-$z$ behavior used in the Fourier transformation, NLO ratio-scheme matching, and a combined continuum and infinite-momentum extrapolation with declared variations. The asymptotic treatment constitutes the main modification relative to the original analysis, as discussed above. Another difference is the ordering of the continuum extrapolation: Ref.~\cite{LatticeParton:2022zqc} performs the $a\to0$ continuum extrapolation at the level of the coordinate-space matrix elements before the Fourier transformation, whereas the present workflow carries the finite-$a$ results through Fourier transformation and matching and performs the continuum and infinite-momentum extrapolations jointly in $x$ space.

\begin{figure*}[t]
  \centering
  \begin{subfigure}[t]{0.488\textwidth}
    \centering
    \includegraphics[width=\linewidth]
      {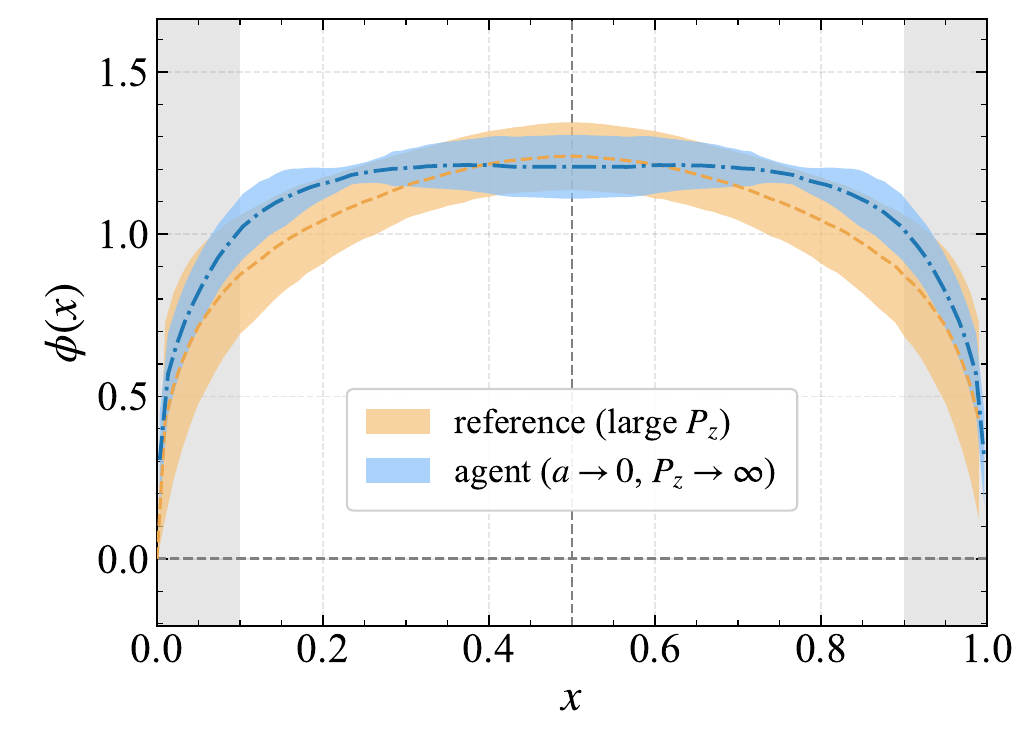}
    \caption{ \centering Pion DA}
    \label{fig:pion_da_compare}
  \end{subfigure}\hfill
  \begin{subfigure}[t]{0.488\textwidth}
    \centering
    \includegraphics[width=\linewidth]
      {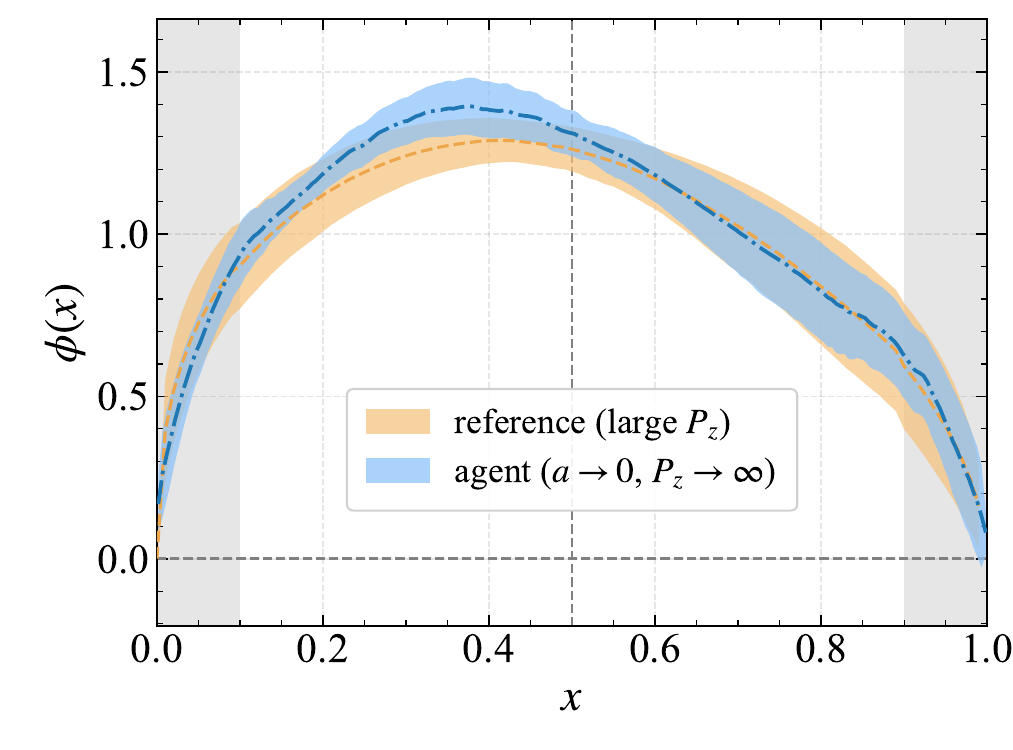}
    \caption{ \centering Kaon DA}
    \label{fig:kaon_da_compare}
  \end{subfigure}
  \caption{Final gauge-invariant comparison of pion and kaon DAs, obtained using the original correlators from Ref.~\cite{LatticeParton:2022zqc}. The blue bands correspond to the \lamet\ joint extrapolation $a\to0$, $P_z\to\infty$, with uncertainties combining the central statistical error and the Fourier-range, matching-scale, and extrapolation-basis components as specified in the manifests. The orange bands represent the results reported in Ref.~\cite{LatticeParton:2022zqc}. The endpoint regions $x<0.1$ and $x>0.9$ are shaded in gray, where LaMET matching suffers large power corrections, hence a quantitative comparison is limited.}
  \label{fig:meson_da_comparison}
\end{figure*}

The two runs differ in how much correlator-level support their selected windows carry. All nine pion selections clear the quality gate, with $\chi^2/\mathrm{dof}=0.100$--$1.346$; for the kaon, the $n_z=6$ selection on the finest ensemble passes with $\chi^2/\mathrm{dof}=0.883$, while the other eight channels, which offer a single candidate window each, miss the gate at a median $\chi^2/\mathrm{dof}$ of $3.54$ and are carried through the documented fallback procedure. Every sample-wise fit completes in both runs, so all downstream artifacts are available. Renormalization then follows the ratio-scheme self-renormalization path at $\mu=2~\mathrm{GeV}$: the common zero-momentum pion reference fixes $z_R(z,a)$ across the three lattice spacings and is combined with the perturbative $\overline{\mathrm{MS}}$ DA coefficients. The same operator-level reference serves the kaon matrix elements, justified by the state independence of ultraviolet renormalization. The Fourier stage evidence-averages the LA and NLA continuations over three prior widths, six candidates wherever both families are admitted, and, following the convention adopted in Ref.~\cite{LatticeParton:2022zqc}, rotates the pion matrix element to the midpoint between the two quark fields and retains its real, $z$-symmetric part, the phase convention in which the pion quasi-DA is real. Both central fits describe the selected long-distance window well. Matching uses the NLO ratio-scheme kernel for the gauge-invariant $\gamma_z\gamma_5$ DA at $\mu=2~\mathrm{GeV}$, with the scale varied by $1/\sqrt2$ and $\sqrt2$. The continuum and infinite-momentum fit is carried out independently at each $x$, so its quality is quoted pointwise and in the region LaMET controls: at $x\simeq0.3$, $0.5$, and $0.7$ the pion returns $\chi^2/\mathrm{dof}=0.34$, $0.43$, and $0.34$, the kaon $5.12$, $5.51$, and $0.78$. Over $0.1\lesssim x\lesssim0.9$ the pion fit stays uniformly acceptable, while the kaon fit is poor over most of that interval, reflecting the quality of its correlator-level input.

\autoref{fig:meson_da_comparison} compares the resulting DAs with Ref.~\cite{LatticeParton:2022zqc}, and \autoref{tab:meson_da_settings} records the resolved setup, the selected windows and ranges, and the fit diagnostics of both runs. For the pion, the uniformly acceptable pointwise fits and the overlap with the reference band through most of $0.1\lesssim x\lesssim0.9$ make this a nontrivial cross-check, and the result retains the expected symmetry about $x=1/2$. The kaon output provides a reproduction consistent with the published result within the propagated systematic uncertainties over the interior region, while the gray endpoint bands indicate the region where a quantitative comparison is not attempted.

The kaon run also gives the most visible exercise of the deterministic consistency check performed at the review stage. The check reports 36 findings, of which nine are errors, and all nine are the same identity mismatch: each renormalized kaon job records the kaon as its consumer hadron, whereas the source renormalization job it consumes records the pion. This is exactly the signature the analysis is designed to produce, since $z_R(z,a)$ is extracted from the pion zero-momentum matrix element and applied to the kaon, an operator-level factor that is state independent for a common discretized operator and lattice action~\cite{LatticeParton:2022zqc}. The check settles identity and leaves the physics to the reader: by naming the edge as an error it requires that justification to be stated explicitly.

\begin{table*}[t]
\caption{
\justifying
Physical setup, agent-selected settings, and fit diagnostics for the gauge-invariant pion- and kaon-DA runs, as reconstructed from the resolved manifests and stage reports. Both use the $96^3\times192$, $64^3\times96$, and $48^3\times64$ ensembles at $a=0.0574$, $0.0882$, and $0.1213~\mathrm{fm}$, respectively. As in \autoref{tab:pion_pdf_settings}, the setup is stated in physical terms and implementation identifiers are shown only for kernels. Systematic variants are propagated through all affected downstream stages.
}
\label{tab:meson_da_settings}
\centering
\footnotesize
\setlength{\tabcolsep}{3pt}
\renewcommand{\arraystretch}{1.18}
\begin{tabular}{@{}p{28mm}p{69mm}p{69mm}@{}}
\hline & Pion DA & Kaon DA \\
\hline
Momenta &
$n_z=6,8,10$; $P_z$=
$(1.35,1.32,1.28)$,
$(1.80,1.76,1.70)$, and
$(2.25,2.20,2.13)~\mathrm{GeV}$
&
same momenta and ensembles
\\
\noalign{\smallskip}

Correlator fit &
one-state fits to the real and imaginary parts of the nonlocal-to-local qDA ratio
&
simultaneous fit to the nonlocal qDA correlator and its aligned $z=0$ two-point correlator
\\

\quad fit window &
$n_z=6:\ [13,18),[8,13),[6,11)$\newline
$n_z=8:\ [9,14),[6,11),[4,9)$\newline
$n_z=10:\ [7,12),[5,8),[3,8)$
&
$n_z=6:\ [12,17),[8,12),[6,11)$\newline
$n_z=8:\ [8,11),[6,9),[4,7)$\newline
$n_z=10:\ [7,12),[5,8),[3,6)$
\\

\quad $\chi^2/\mathrm{dof}$
&
$n_z=6:\ (0.251,0.130,0.422)$\newline
$n_z=8:\ (0.873,0.140,0.500)$\newline
$n_z=10:\ (0.777,1.346,0.100)$
&
$n_z=6:\ (1.093,4.198,3.141)$\newline
$n_z=8:\ (5.141,9.516,9.540)$\newline
$n_z=10:\ (3.231,13.313,65.299)$
\\
\noalign{\smallskip}
\hline
\noalign{\smallskip}

Renormalization
&
ratio scheme with self-renormalization; the common zero-momentum pion reference determines $z_R(z,a)$; $\mu=2~\mathrm{GeV}$
&
same reference and strategy applied to the kaon matrix elements
\\

\noalign{\smallskip}
\hline
\noalign{\smallskip}

Asymptotic region &
$n_z=6:\ (0.52\text{--}1.21),(0.53\text{--}1.50),(0.61\text{--}1.46)$\newline
$n_z=8:\ (0.52\text{--}1.49),(0.53\text{--}1.50),(0.61\text{--}1.46)$\newline
$n_z=10:\ (0.52\text{--}1.49),(0.53\text{--}1.50),(0.61\text{--}1.46)$
&
$n_z=6:\ (0.75\text{--}1.49),(0.62\text{--}1.50),(0.97\text{--}1.21)$\newline
$n_z=8:\ (0.75\text{--}1.26),(0.62\text{--}0.97),(0.73\text{--}0.97)$\newline
$n_z=10:\ (0.75\text{--}1.21),(0.79\text{--}0.97),(0.85\text{--}1.09)$
\\

\quad tail model &
LA/NLA continuation with \texttt{linear}; \newline
$\chi^2/\mathrm{dof}=0.008$--$0.256$
&
same continuation; $\chi^2/\mathrm{dof}=10^{-4}$--$0.218$
\\
\noalign{\smallskip}
\hline
\noalign{\smallskip}

Matching
&
NLO matching at $\mu=2~\mathrm{GeV}$
&
NLO matching at $\mu=2~\mathrm{GeV}$
\\
\noalign{\smallskip}

Continuum and infinite-momentum fit
&
pointwise simultaneous fit with $a^2$, $P_z^{-2}$, and $(aP_z)^2$ corrections
&
$x$-independent $a^2$ correction with pointwise $P_z^{-2}$, $(aP_z)^2$, and $P_z^{-4}$ corrections
\\

\quad fit quality\newline
at $x\simeq0.3,0.5,0.7$
&
$\chi^2/\mathrm{dof}=(0.34,0.43,0.34)$
&
$\chi^2/\mathrm{dof}=(5.12,5.51,0.78)$
\\
\noalign{\smallskip}

Systematic variations
&
Fourier onset shifted by $\pm a$; matching scale varied by $1/\sqrt{2}$ and $\sqrt{2}$; separate additions of $a^4$, $P_z^{-4}$, and $(aP_z)^4$ corrections
&
one-step earlier Fourier onset; the same scale variations; separate addition of $a^4$, removal of $P_z^{-4}$, and addition of $(aP_z)^4$ corrections
\\

\hline
\end{tabular}
\end{table*}

\section{Status and Outlook}
\label{sec:outlook}

We have introduced \lamet, an LLM-agent framework that represents a LaMET lattice-QCD analysis as a declarative job-DAG manifest. Plan Mode prepares and validates the specification, while a manifest that already satisfies the relevant contracts proceeds directly to execution. Execution applies the declared subset of six stage contracts through deterministic numerical workflows, with data-dependent choices restricted to bounded, schema-validated \texttt{ask} operations. Correlator analysis allows the greatest flexibility; renormalization branches on a user-declared scheme; and the subsequent stages follow increasingly constrained workflows. The terminal review is read-only and synthesizes recorded artifacts without modifying them. This interface makes the boundary between model judgment and numerical execution explicit, permits user intervention, and supports partial workflows initialized from external artifacts. The framework is released as an open-source Python package with a \texttt{lamet-agent} console entry point providing \texttt{validate}, \texttt{plan}, and \texttt{run} subcommands.

The four analyses of \autoref{sec:case_study} are the present validation: pion PDFs in the gauge-invariant and Coulomb-gauge formulations at a single lattice spacing, and pion and kaon DAs across three spacings with continuum and infinite-momentum extrapolation. Together, they exercise two observables, two operator constructions, two renormalization schemes, and both single-ensemble and extrapolated workflows, and they reproduce the published results. Because these four cases are representative along each of these dimensions, the present studies validate the execution of expert-specified workflows across a broad range of LaMET analyses.

Several validation questions therefore remain open. The systematic-budget tools are implemented, but their ansatz families and uncertainty coverage need testing beyond the variants exercised here. Comparisons among LLM providers, against human-expert baselines, and across repeated runs are needed to characterize the stability and cost of agent decisions, and human-in-the-loop studies should identify which choices can be delegated safely and which should require explicit approval. The terminal review remains a secondary synthesis: numerical claims are authoritative only when grounded in the stage reports and NetCDF summaries.

Several extensions are already planned. On the observable side, workflow contracts for TMDs and generalized TMDs (GTMDs)~\cite{Ji:2018hvs,Musch:2010ka} will add transverse structure through extended renormalization, Fourier, and matching paths, without changing the manifest or job-DAG architecture. Threshold resummation of the matching kernels will extend the current leading-renormalon-resummed treatment. Within correlator analysis, the candidate space will be broadened from fixed-state fits to variational methods based on the generalized eigenvalue problem~\cite{Blossier:2009kd}, making multi-state resolution itself a selectable choice, and the stage will be generalized beyond quasi-distribution matrix elements to current-current correlators following the large-momentum expansion of Ref.~\cite{Zhang:2026lle}. Each extension enters as a new stage contract, \texttt{ask} specification, or kernel within the same declarative interface, preserving the separation between deterministic physics operations and constrained, inspectable model judgment.

More broadly, a constrained agentic workflow can present a theoretical framework as more than a collection of equations and disconnected implementations. Combining the analysis pipeline, expert guidance, deterministic tools, and inspectable provenance makes the methodology executable and extensible. Such automation can reduce duplicated implementation effort, lower the barrier to reproducible use, and provide a common foundation for testing and extending first-principles analysis workflows. This may point toward a new paradigm in which theoretical frameworks are introduced through executable, reproducible agentic systems.

\begin{acknowledgments}
We thank Jianhui Zhang, Rui Zhang, Peter Petreczky, Xiang Gao, Yong Zhao and Jun Hua for useful discussions during the development of \lamet.
We are especially grateful to the ANL--BNL collaboration and the Lattice Parton Collaboration (LPC) for making the numerical data associated with their respective publications publicly available. The availability of these open data was essential for the direct validation and reproduction studies carried out in this work. Information on accessing the open data used in all four examples is provided in the GitHub repository at \url{https://github.com/AI4LGT/lamet-agent}.
Code development was performed with substantial assistance from Cursor, OpenAI's Codex, and Anthropic's Claude Code.
This manuscript has been authored by Fermi Forward Discovery Group, LLC under Contract No. 89243024CSC000002 with the U.S. Department of Energy, Office of Science, Office of High Energy Physics.
This material is based upon work supported by the U.S. Department of Energy, Office of Science, Office of Nuclear Physics through Contract No.~DE-SC0012704, No.~DE-AC02-06CH11357, within the framework of Scientific Discovery through Advanced Computing (SciDAC) award Fundamental Nuclear Physics at the Exascale and Beyond, and under the umbrella of the Quark-Gluon Tomography (QGT) Topical Collaboration with Award DE-SC0023646. {This work is partially supported by the AmSC project QCD for Foundational Model.} This material is based upon work supported by the National Science Foundation under Grant No. 2323116 for the Construction for the Leadership Class Computing Facility (LCCF), and specifically through the sub-award UTAUS-SUB00001853 managed by the Texas Advanced Computing Center (TACC) at The University of Texas at Austin. This material is based upon work supported by the U.S. Department of Energy, Office of Science, Office of Nuclear Physics through Contract No. DE-SC0012704, and within the framework of Scientific Discovery through Advanced Computing (SciDAC) award Fundamental Nuclear Physics at the Exascale and Beyond. This work is supported in part by NSFC grant No. 12375080, the Ministry of Science and Technology of China under Grant No. 2024YFA1611004, and by CUHK-Shenzhen under grant No. UDF01002851. 
\end{acknowledgments}

\bibliography{main}

\clearpage
\appendix
\begin{widetext}
\section{Parameters Manual}
\label{app:parameters_manual}
This appendix is a manifest reference for the current stage contracts. Only JSON fields that may be authored in a manifest are listed. Stage parameters may be supplied in a stage-level \texttt{defaults} object or directly in an individual job, with the job value taking precedence; role-named dependencies are always placed under \texttt{jobs[].inputs}. The final column condenses the corresponding \texttt{physics} explanations in \texttt{manifest.py}, the base contract, or the relevant stage \texttt{contract.py}; where several rules govern one field, it retains the principal physical action and activation condition.

\subsection{Metadata}
The \texttt{metadata} object defines the run identity, directory, physical parameters, and the common execution and resampling context inherited by all jobs. Observable and parton labels serve to activate conditional stage contracts, while the sampling fields govern the binning of raw configurations and the propagation of uncertainty replicas. Stage-specific fitting, renormalization, and matching choices are explicitly excluded from this object. The complete list of accepted parameters is provided in \autoref{tab:app_metadata_params}.

\begin{table*}[th!]
\caption{Manifest fields accepted in the \texttt{metadata} object.}
\label{tab:app_metadata_params}
\centering
{\scriptsize
\setlength{\tabcolsep}{4pt}
\renewcommand{\arraystretch}{1.08}
\begin{tabular}{p{48mm}p{43mm}p{64mm}}
\hline
JSON field & Accepted value & Physical meaning or condition \\
\hline
\texttt{run\_id} & string & Human-readable identifier for one analysis run and its artifacts. \\
\texttt{root\_directory} & path string & Filesystem base used to resolve relative input paths. \\
\texttt{artifacts\_directory} & path string & Directory receiving per-job outputs and diagnostics. \\
\texttt{random\_seed} & nonnegative integer & Common seed for reproducible resampling and stochastic numerical procedures. \\
\texttt{workers} & positive integer & Run-wide limit on concurrent sample-wise calculations. \\
\texttt{target\_observable} & \texttt{pdf}, \texttt{da}, or \texttt{gpd} & Physical distribution produced by the workflow; it activates observable-specific stage fields. \\
\texttt{parton} & \texttt{quark}; default \texttt{quark} & Selects quark correlators and quark matching conventions; gluon paths are not yet released. \\
\texttt{resample\_mode} & \texttt{jackknife} or \texttt{bootstrap} & Jackknife leaves one group out at a time; bootstrap resamples with replacement to generate uncertainty replicas. \\
\texttt{sample\_error\_mode} & \texttt{covariance}, \texttt{variance}, or \texttt{one\_sigma} & Retains covariance, reports variance-only errors, or uses the bootstrap median and 16/84-percentile interval. \\
\texttt{parameter\_recommendation\_retries} & nonnegative integer; default 1 & Bounded number of extra LLM proposals after an invalid per-job recommendation. \\
\texttt{samples} & positive integer & Number of bootstrap replicas used for uncertainty estimation; bootstrap only. \\
\texttt{bin\_size} & positive integer & Number of raw configurations combined before resampling. \\
\hline
\end{tabular}}
\end{table*}

\subsection{Stages}
The ordered \texttt{stages} mapping, together with globally unique job identifiers, defines the analysis DAG. Stage-level defaults supply common parameter choices, which individual jobs could override as needed. Role-named inputs explicitly declare every upstream dependency and external artifact, ensuring full traceability. This design accommodates both complete pipelines and partial workflows that resume from a declared intermediate file. The general structure is presented in \autoref{tab:app_stage_params}, followed by the stage-specific job fields.

\begin{table*}[th!]
\caption{Generic manifest fields accepted for every numerical or review stage.}
\label{tab:app_stage_params}
\centering
{\scriptsize
\setlength{\tabcolsep}{4pt}
\renewcommand{\arraystretch}{1.08}
\begin{tabular}{p{48mm}p{43mm}p{64mm}}
\hline
JSON field & Accepted value & Physical meaning or condition \\
\hline
\texttt{stages} & nonempty ordered object & Ordered stage mapping defining the sequence of physical transformations. \\
\texttt{stages.<stage>.defaults} & object; optional & Authored common values inherited by jobs in that stage. \\
\texttt{stages.<stage>.jobs} & nonempty ordered list & Physical jobs retain authored order before systematic expansion. \\
\texttt{jobs[].id} & safe identifier string & Stable graph identifier used for dependencies and job artifacts. \\
\texttt{jobs[].inputs} & role-to-source object; default \texttt{\{\}} & Maps each physical input role to an upstream job, external file, or explicitly permitted source form. \\
\texttt{jobs[].<stage parameter>} & stage-specific value & Job-specific physical choice overriding the inherited stage default. \\
\hline
\end{tabular}}
\end{table*}

\subsubsection{Correlator Analysis}
\label{app:correlator_stage}
Correlator analysis converts selected two- and three-point correlation functions, or a nonlocal-to-local qDA ratio, into resampled spectra or bare matrix elements. The fit scope alone fixes both the physical observable being isolated and the way spectral information enters: atoms joined by \texttt{+} inside one entry share a correlated joint likelihood, successive list entries chain the preceding posterior forward, and omitting the \texttt{2pt} atom leaves an independent ratio fit with no separate two-point model. For least-squares analysis, candidate windows and models are assessed on the sample-average data before one choice is held fixed across resamples; the Lanczos path instead controls the nested-bootstrap Ritz extraction. These choices directly govern excited-state contamination and numerical stability, so their selected values and fit quality must be read together. The accepted fields are listed in \autoref{tab:app_correlator_params}.

\begin{table*}[th!]
\caption{Manifest fields accepted by \texttt{correlator\_analysis}. Lsqfit- and Lanczos-specific fields are active only for the corresponding \texttt{analysis\_method}.}
\label{tab:app_correlator_params}
\centering
{\scriptsize
\setlength{\tabcolsep}{4pt}
\renewcommand{\arraystretch}{1.05}
\begin{tabular}{p{48mm}p{48mm}p{59mm}}
\hline
JSON field & Accepted value & Physical meaning or condition \\
\hline
\texttt{inputs.correlators} & nonempty list of descriptor records & Explicit ordered correlator selection; one descriptor may contain many records. \\
\texttt{inputs.correlators[].json} & string ending in \texttt{.json} & Descriptor containing the selected correlator's data definition. \\
\texttt{inputs.correlators[].id} & nonempty string & Unique record identifier selecting one correlator from that descriptor. \\
\texttt{analysis\_method} & \texttt{lsqfit} or \texttt{lanczos}; default \texttt{lsqfit} & Uses sample-wise nonlinear least squares or a Lanczos extraction. \\
\texttt{component} & \texttt{re}, \texttt{im}, or \texttt{both} & Selects the real, imaginary, or both matrix-element channels. \\
\texttt{nstate} (\texttt{lsqfit}) & object mapping each \texttt{fit\_scope} atom to a nonempty list of positive integers & Per-correlator spectral truncations scanned by the fit; keys must be bare atoms, never joint \texttt{+} strings. \\
\texttt{nstate} (\texttt{lanczos}) & list holding one positive integer & Single authored count of exported Ritz states; Lanczos infers its internal order. \\
\texttt{fit\_scope} & nonempty ordered list of \texttt{+}-joined atoms drawn from \texttt{2pt}, \texttt{3pt}, \texttt{qda}, \texttt{FH}, \texttt{3pt\_ratio}, \texttt{qda\_ratio} & Each entry is one joint likelihood and list order denotes chained posterior propagation; \texttt{2pt} may appear once and only in the first entry, qDA atoms cannot mix with three-point/FH atoms, and raw and ratio forms of one correlator are mutually exclusive. \\
\texttt{fitting\_form} & \texttt{Breit} or \texttt{NonBreit} & Breit: $|\textbf{p}_i|=|\mathbf{p}_f|$; NonBreit: otherwise. NonBreit requires a \texttt{3pt} or \texttt{3pt\_ratio} atom and admits only an accompanying \texttt{2pt}; qDA requires Breit. \\
\texttt{prior\_width} & nonempty list of positive floats; default \texttt{[1.0]} & Gaussian prior scales scanned to test sensitivity of underconstrained fit directions. \\
\texttt{model\_average} & Boolean & false: publishes the single best-fit candidate; true: Bayesian-model average over all candidates. \\
\texttt{pt2\_windows[].tmin}, \texttt{tmax} & nonnegative and positive integer with $t_{\min}<t_{\max}$ & Candidate Euclidean 2pt intervals $[t_{\min},t_{\max})$ used to test fit-range stability. \\
\texttt{pt3\_windows[].tsep\_ls} & nonempty list of unique positive integers & Source--sink separations combined in one three-point candidate. \\
\texttt{pt3\_windows[].tau\_cut} & nonnegative integer, unique across windows & Number of insertion-time slices removed near each endpoint; at least one insertion remains. \\
\texttt{svdcut} & positive number; default $10^{-12}$ & Relative covariance singular-value cutoff suppressing numerically unresolved fit directions. \\
\texttt{posterior\_prior\_error\_scale} & positive number & Multiplier widening propagated posterior widths in chained and sample-wise fits. \\
\texttt{q\_min} & number in $[0,1]$ & Preferred fit-quality threshold $Q\ge q_{\min}$; a finite-$Q$ fallback is allowed after recommendation retries. \\
\texttt{scope} & \texttt{2pt\_spectrum} or \texttt{3pt\_matrix} & Selects a two-point spectrum or three-point matrix element for Lanczos analysis. \\
\texttt{inner\_samples} & positive integer; default 200 & Size of the inner bootstrap ensemble drawn for each outer sample in Cullum--Willoughby filtering and median aggregation. \\
\texttt{precision} & nonnegative integer; default 0 & Decimal precision of the Lanczos recurrence; zero selects NumPy double precision. \\
\texttt{final\_iteration} & positive integer; default none & Lanczos iteration published as the three-point matrix; omitting it selects the second-to-last usable iteration. \\
\hline
\end{tabular}}
\end{table*}

\subsubsection{Renormalization}
\label{app:renormalization_stage}
Renormalization separates how a factor is obtained from the scheme in which it is applied. An external-denominator job consumes a supplied factor directly, whereas self-renormalization first fits a reusable $z_R(z,a)$ from reference matrix elements and then applies it to one or more targets. The ratio, $\overline{\mathrm{MS}}$, and hybrid schemes activate different conversion, linear-mass, and switching-distance terms, so the strategy, operation type, scheme, and input roles form one conditional contract. The accepted fields are listed in \autoref{tab:app_renorm_params}.

\begin{table*}[th!]
\caption{Manifest fields accepted by \texttt{renormalization}. The selected \texttt{strategy}, \texttt{type}, and \texttt{scheme} jointly determine
which conditional fields and input roles are required.}
\label{tab:app_renorm_params}
\centering
{\scriptsize
\setlength{\tabcolsep}{4pt}
\renewcommand{\arraystretch}{1.05}
\begin{tabular}{p{49mm}p{45mm}p{61mm}}
\hline
JSON field & Accepted value & Physical meaning or condition \\
\hline
\texttt{strategy} & \texttt{external\_denominator}, \texttt{self\_renormalization} & Uses a factor determined outside the workflow or extracts a reusable factor from reference matrix elements. \\
\texttt{normalization} & Boolean & If true, divides every sample by its unique $z=0$ value before renormalization. \\
\texttt{scheme} & \texttt{ratio}, \texttt{hybrid}, or \texttt{msbar} & Selects how the supplied or fitted factor enters; hybrid separates short- and long-distance prescriptions. \\
\texttt{type} & \texttt{apply} for \texttt{external\_denominator}, \texttt{fit}/\texttt{apply} for \texttt{self\_renormalization} & Fits a reusable $z_R$ from references or applies the selected prescription to a target; external factors are apply-only. \\
\texttt{inputs.target} & job id or file source & Bare matrix element to be renormalized. \\
\texttt{inputs.denominator} & job id or file source; a nonzero constant only where permitted & External factor dividing the target; hybrid use requires a coordinate-dependent source and also anchors continuity in self-hybrid application. \\
\texttt{inputs.reference} & job/file source or ordered source list & Reference $(a,z)$ data, possibly at several lattice spacings, used to fit the reusable factor. \\
\texttt{inputs.zR} & job id or file source & Reusable factor produced by a self-renormalization fit and applied to the target. \\
\texttt{zs\_fm} & positive number in fm & Hybrid switching distance between short- and long-distance treatments; it must lie on the $z$ grid. \\
\texttt{m0\_gev} & finite number in GeV & Renormalon-related linear mass term for external hybrid renormalization or target-operator remapping. \\
\texttt{delta\_m\_gev} & finite number in GeV & Wilson-line linear-power-divergence mass term; external-denominator hybrid only. \\
\texttt{mu} & positive number & Perturbative scale for $\overline{\mathrm{MS}}$ scheme and self-renormalization provenance. \\
\texttt{kernel\_id} & public renormalization-kernel id & Operator- and channel-specific coordinate-space $\overline{\mathrm{MS}}$ scheme formula and perturbative order; self-renormalization only. \\
\texttt{kernel\_parameters} & object; default \texttt{\{\}} & Non-coordinate overrides validated against the selected kernel signature; $z$ is supplied by the data and cannot be overridden. \\
\texttt{svdcut} & positive number; default $10^{-12}$ & Relative covariance singular-value cutoff regularizing the correlated self-renormalization fit. \\
\texttt{LambdaQCD\_gev} & positive number in GeV & Common $\Lambda_{\rm QCD}$ governing the ultraviolet running in the self-renormalization fit and its application. \\
\texttt{z\_coverage\_policy} & \texttt{strict}, \texttt{intersection}, or \texttt{extrapolate}; default \texttt{extrapolate} & Requires full target coverage, keeps only the overlap, or extrapolates $z_R$ only toward larger $|z|$. \\
\texttt{d} & finite number & Reference-operator finite correction in a fit, or target-operator finite-term remapping in an application. \\
\hline
\end{tabular}}
\end{table*}

\subsubsection{Fourier Transform}
\label{app:fourier_stage}
Fourier transformation maps a finite, resampled coordinate-space renormalized matrix element to a quasi-distribution. Its physical choices specify the measured interval used to constrain the large-distance continuation, the completion of the negative-$z$ branch, and any DA or GPD phase convention imposed before the transform. Range and tail-model candidates are assessed on the sample-average data; the selected range is then held fixed. When model averaging is off, the selected tail model is likewise held fixed and applied to every resample; when it is on, every authored model is fitted on that frozen range and the resampled results are combined by evidence-weighted means. The output $x$ grid controls the numerical representation. The accepted fields are listed in \autoref{tab:app_fourier_params}.

\begin{table*}[th!]
\caption{Manifest fields accepted by \texttt{fourier\_transform}.}
\label{tab:app_fourier_params}
\centering
{\scriptsize
\setlength{\tabcolsep}{4pt}
\renewcommand{\arraystretch}{1.05}
\begin{tabular}{p{50mm}p{47mm}p{58mm}}
\hline
JSON field & Accepted value & Physical meaning or condition \\
\hline
\texttt{inputs.input} & job id or file source & Renormalized coordinate-space matrix element. \\
\texttt{inputs.hermitian\_partner} & job id or file source; default none & Exchanged-flow GPD partner with reversed source and sink momenta, supplying the negative-$z$ branch by Hermiticity. \\
\texttt{quasi\_y\_ls} & increasing list or \texttt{\{start,stop,num\}} & Dimensionless $y$ grid. \\
\texttt{zmin\_fm}, \texttt{zmax\_fm} & nonempty lists of nonnegative and positive numbers & Candidate lower/upper tail-fit boundaries; one common range is selected from sample-average data before resample fits. \\
\texttt{tail\_window\_step\_offset} & integer; default 0 & Shifts every $z_{\min}$ by the signed distance $n a_s$ while holding $z_{\max}$ fixed; zero is the unshifted central job. \\
\texttt{zmax\_ext\_fm} & positive number & Requested maximum $|z|$ in the finite sum. \\
\texttt{smooth} & \texttt{linear} or \texttt{none} & Linear blends data to tail over $[z_{\min},z_{\max}]$; none keeps unit data weight through $z_{\rm ext}$, with tail weight only beyond it. \\
\texttt{scheme\_scan.order} & nonempty list of \texttt{LA}, \texttt{NLA} & LA retains the leading large-$|z|$ form; NLA adds the next inverse-distance term. \\
\texttt{scheme\_scan.sector} & \texttt{sea}, \texttt{valence}, \texttt{singlet}, or \texttt{full} & Forms the sea, quark-minus-antiquark, or quark-plus-antiquark combination, or the complete complex result; \texttt{sea} is GPD-only and DA requires \texttt{full}. \\
\texttt{scheme\_scan.Lambda0\_gev} & nonnegative number in GeV & Fixed infrared offset in $\exp[-(\Lambda+\Lambda_0)|z|]$, with $\Lambda$ fitted. \\
\shortstack[l]{\texttt{scheme\_scan.posterior\_prior\_}\\
\texttt{error\_scale}} & nonempty list of positive numbers & Multipliers of sample-average posterior widths defining resample priors and regularization variants. \\
\texttt{scheme\_scan.model\_average} & Boolean & false: publishes the sample-average-selected model for every resample; true: evidence-weighted means over valid models. \\
\texttt{scheme\_scan.q\_min} & number in $[0,1]$; default 0.05 & Preferred $Q$ threshold for range and non-averaged model selection, with maximum-$Q$ fallback. \\
\texttt{scheme\_scan.max\_schemes} & positive integer; default 200 & Cost cap on the model--$z_{\min}$--$z_{\max}$ candidates used for sample-average range selection; it bounds no formula. \\
\texttt{phase\_transfer\_da} & Boolean; DA only & Rotates by $e^{i P_z z/2}$, projects the real midpoint-symmetric channel, and rotates back; false retains complex samples. \\
\texttt{psi1\_flavor\_class}, \texttt{psi2\_flavor\_class} & \texttt{light} or \texttt{heavy}; DA only & Ordered endpoint mass classes selecting the charge-conjugation and asymmetric-tail constraints. \\
\texttt{phase\_transfer\_gpd} & \texttt{mid\_at\_0}, \texttt{barpsi\_at\_0}, or \texttt{psi\_at\_0}; default \texttt{mid\_at\_0} & Applies opposite half-transfer phases, no transfer, or conjugated endpoint exchange before GPD Hermitian completion. \\
\hline
\end{tabular}}
\end{table*}

\subsubsection{Perturbative Matching}
\label{app:matching_stage}
Perturbative matching applies a deterministic matching kernel to convert a finite-momentum quasi-distribution into its light-cone counterpart. The kernel is not named in the manifest: its identity is derived at runtime from the perturbative order and resummation authored here together with the quasi-input provenance, namely the parton, observable, gauge construction, operator, renormalization scheme, and, for running-coupling resummation, the source component. The renormalization scheme and the hybrid switching distance are likewise read from the upstream attributes, so an inconsistent choice cannot be expressed. The scale and output interval specify the factorization setup. Because this stage is not a fit, its validation rests on convention, grid, and provenance consistency. The accepted fields are listed in \autoref{tab:app_matching_params}.

\begin{table*}[th!]
\caption{Manifest fields accepted by \texttt{perturbative\_matching}. The kernel filename is derived from these fields and the quasi-input provenance.}
\label{tab:app_matching_params}
\centering
{\scriptsize
\setlength{\tabcolsep}{4pt}
\renewcommand{\arraystretch}{1.08}
\begin{tabular}{p{49mm}p{48mm}p{58mm}}
\hline
JSON field & Accepted value & Physical meaning or condition \\
\hline
\texttt{inputs.quasi} & job id or file source & One quasi-distribution whose attributes supply $P_z$, the hybrid $z_s$, and the parton, observable, gauge, operator, scheme, and component provenance fixing the kernel. \\
\texttt{order} & \texttt{nlo} & Only next-to-leading-order matching kernels are currently available. \\
\texttt{resummation} & \texttt{""}, \texttt{rgr}, or \texttt{lrr}; default \texttt{""} & Empty selects fixed order; \texttt{rgr} resums the running coupling and takes its real/imaginary kernel suffix from the upstream source component; \texttt{lrr} resums the leading renormalon. \\
\texttt{mu} & positive number in GeV & $\overline{\mathrm{MS}}$ light-cone scale entering the coefficient function through $\ln(4y^2P_z^2/\mu^2)$. \\
\texttt{lc\_x\_ls} & increasing list or \texttt{\{start,stop\}} & Exact output grid, or a closed window selecting existing quasi-grid points without interpolation. \\
\texttt{kernel\_parameters} & object; default \texttt{\{\}} & Overrides checked against the selected kernel signature; every kernel accepts the plus-prescription regulator \texttt{eps}, and RGR kernels add \texttt{kappa} and \texttt{mu\_min\_gev}, which set $x_{\min}=\mu_{\min}/(2\kappa P_z)$ and keep $\mu_0$ above the Landau pole. The grids and $z_s$ cannot be overridden. \\
\hline
\end{tabular}}
\end{table*}

\subsubsection{Extrapolation}
\label{app:extrapolation_stage}
The extrapolation procedure fits a set of matched distributions to obtain $h_0(x)$ in the limit of vanishing lattice artifacts, finite-momentum effects, unphysical quark-mass effects, and finite-volume corrections. The correction coefficients may be either shared globally across the entire $x$ grid or fitted independently at each grid point; this distinction separates a global scaling effect from a point-dependent shape distortion. A separate budget-combination step assembles the explicitly grouped central and variant outputs into pointwise statistical and systematic uncertainties without performing any refitting. A physically meaningful interpretation of the final budget, however, requires that the central fit be acceptable. The list of accepted parameters for both the extrapolation and the budget operation is given in \autoref{tab:app_extrapolation_params}.

\begin{table*}[tbp]
\caption{Manifest fields accepted by \texttt{extrapolation}. Fit-only and budget-only fields are selected by \texttt{operation}.}
\label{tab:app_extrapolation_params}
\centering
{\scriptsize
\setlength{\tabcolsep}{4pt}
\renewcommand{\arraystretch}{1.05}
\begin{tabular}{p{51mm}p{49mm}p{55mm}}
\hline
JSON field & Accepted value & Physical meaning or condition \\
\hline
\texttt{operation} & \texttt{fit} or \texttt{systematics\_budget}; default \texttt{fit} & Fits $h_0(x)$ where the selected correction bases vanish, or performs no refit and assembles statistical and systematic errors. \\
\texttt{inputs.distributions} & nonempty ordered list of job ids/file sources & Matched finite-parameter distributions for the scaling fit, or indexed central and variant distributions for the budget. \\
\texttt{x\_independent\_terms} & unique list; default \texttt{[]} & Bases in $h=h_0(x)+\sum_t c_tB_t$ with one correction coefficient shared over all $x$. \\
\texttt{x\_dependent\_terms} & unique list; default \texttt{[]} & Bases with independent $c_t(x)$, allowing cutoff, momentum, mass, or volume effects to distort the shape; disjoint from the shared list. \\
Supported term identifiers & \texttt{a}, \texttt{a2}, \texttt{a4}, \texttt{ap2}, \texttt{ap4}, \texttt{exp\_mpi\_L}, \texttt{exp\_sqrt2\_mpi\_L}, \texttt{mpi2}, \texttt{mpi4\_log\_mpi2}, \texttt{inv\_p2}, \texttt{inv\_p4} & Implement $a_s^{1,2,4}$, raw $(a_sP)^{2,4}$, finite-volume exponentials, physical-point-subtracted pion-mass terms, and higher-twist terms $P^{-2,-4}$. \\
\texttt{priors} & exactly \texttt{\{mean,sdev\}}; default \texttt{\{0,3\}} & Common numerical Gaussian prior for $h_0$ and all correction coefficients in their native units. \\
\texttt{x\_covariance} & Boolean; default \texttt{false} & Retains fixed-$x$ covariance by compatible resample group, and optionally cross-$x$ covariance within each group. \\
\texttt{pdep\_gev} & nonempty list of unique positive numbers & Momenta for post-fit $h_0+c_{P^{-2}}/P^2+c_{P^{-4}}/P^4$ diagnostic curves only; they add no fit data. \\
\texttt{physical\_pion\_mass\_gev} & positive number; default 0.135 & Chiral target at which the implemented pion-mass correction bases vanish. \\
\texttt{posterior\_prior\_error\_scale} & positive number & Multiplier of sample-average posterior widths used as priors for each resample fit. \\
\texttt{systematics\_prescription} & \texttt{variant\_envelope\_quadrature}; default as shown & Uses variant central values only: $|\mathrm{variant}-\mathrm{main}|$ for one or max--min for several, then combines independent components and the main statistical error in quadrature. \\
\texttt{systematics\_groups} & object with \texttt{main}, \texttt{zs}, \texttt{lambda\_extrapolation}, \texttt{lamet\_scale}, \texttt{other\_extrapolations} & Partitions every ordered input once into authored uncertainty components. \\
\hline
\end{tabular}}
\end{table*}

\subsubsection{Review}
\label{app:review_stage}
Review is a terminal, evidence-bounded synthesis. Its ordered inputs define the results available to the reviewer, and the requested checks deterministically test selected cross-stage identities, units, kinematics, schemes, grids, resampling, and extrapolation metadata. An optional catalog may add a bounded number of papers as methodological context, but literature cannot supply a missing run value or override a stage diagnostic. The review writes recommendations without changing the manifest or any numerical artifact. The accepted fields are listed in \autoref{tab:app_review_params}.

\begin{table*}[th!]
\caption{Manifest fields accepted by \texttt{review}.}
\label{tab:app_review_params}
\centering
{\scriptsize
\setlength{\tabcolsep}{4pt}
\renewcommand{\arraystretch}{1.08}
\begin{tabular}{p{48mm}p{48mm}p{59mm}}
\hline
JSON field & Accepted value & Physical meaning or condition \\
\hline
\texttt{inputs.results} & nonempty ordered list of job ids/file sources & Explicit, ordered prior results defining the evidence available to review. \\
\texttt{catalog} & string & Selects the built-in catalog or a literature catalog resolved relative to the run root. \\
\texttt{max\_papers} & positive integer & Bounds the number of selected full-text papers exposed to the reviewer. \\
\texttt{report\_language} & \texttt{en} or \texttt{ch} & Generates the review prose directly in English or Chinese. \\
\texttt{checks} & nonempty list drawn from \texttt{identity}, \texttt{units}, \texttt{kinematics}, \texttt{schemes}, \texttt{grids}, \texttt{resampling}, \texttt{extrapolation} & Explicit controlled set of cross-result consistency checks requested from review. \\
\hline
\end{tabular}}
\end{table*}

\subsection{Systematics}
The \texttt{systematics} object is designed to quantify systematic uncertainties by defining a set of alternative analysis variants that branch off from the nominal central job graph. Unlike an error model inferred at runtime, these variants are explicitly specified at configuration time. Each variant modifies exactly one supported parameter at the stage where the corresponding systematic effect enters, and its manifest expansion propagates that modification through all downstream jobs. A subsequent extrapolation-budget job must explicitly group the resulting branches and then combine their pointwise differences to yield the final systematic uncertainty estimates. The currently supported variants for Fourier transformation, perturbative matching, and extrapolation are enumerated in \autoref{tab:app_systematics_params}.

\begin{table*}[th!]
\caption{Manifest fields accepted under the \texttt{systematics} object. No stage-local systematics contract is currently defined for correlator analysis or renormalization.}
\label{tab:app_systematics_params}
\centering
{\scriptsize
\setlength{\tabcolsep}{4pt}
\renewcommand{\arraystretch}{1.08}
\begin{tabular}{p{53mm}p{46mm}p{56mm}}
\hline
JSON field & Accepted value & Physical meaning or condition \\
\hline
\texttt{systematics.<stage>} & object keyed by \texttt{fourier\_transform}, \texttt{perturbative\_matching}, or \texttt{extrapolation} & Adds uncertainty variants only to a stage already present in the central workflow. \\
\texttt{defaults} & object; default \texttt{\{\}} & Common values inherited by variants before explicit overrides; central jobs are unchanged. \\
\texttt{fourier\_transform.variants[].id} & safe lowercase identifier & Stable suffix and provenance label for one Fourier-tail variation. \\
\shortstack[l]{\texttt{fourier\_transform.variants[].}\\
\texttt{tail\_window\_step\_offset}} & nonzero integer & Moves every $z_{\min}$ by the signed distance $n a_s$, leaving $z_{\max}$ and other Fourier settings fixed. \\
\texttt{perturbative\_matching.variants[].id} & safe lowercase identifier & Stable suffix and provenance label for one matching-scale variation. \\
\shortstack[l]{\texttt{perturbative\_matching.variants[].}\\
\texttt{mu\_factor}} & positive number other than 1 & Multiplies the central $\mu$, conventionally by $1/\sqrt2$ or $\sqrt2$, for the matching-scale uncertainty. \\
\texttt{extrapolation.variants[].id} & safe lowercase identifier & Stable suffix and uncertainty-provenance label for an alternative extrapolation ansatz. \\
\shortstack[l]{\texttt{append\_x\_independent\_terms},\\
\texttt{remove\_x\_independent\_terms}} & lists of supported basis identifiers; default \texttt{[]} & Add or remove corrections whose coefficient is shared over all $x$, testing a shape-independent scaling assumption. \\
\shortstack[l]{\texttt{append\_x\_dependent\_terms},\\
\texttt{remove\_x\_dependent\_terms}} & lists of supported basis identifiers; default \texttt{[]} & Add or remove corrections with independent coefficients $c_t(x)$, testing an $x$-dependent distortion. \\
\hline
\end{tabular}}
\end{table*}

\end{widetext}

\end{document}